\documentclass[11pt,preprint]{emulateapj}
\usepackage{graphicx}
\usepackage{natbib}
\usepackage{amsmath}
\usepackage{multirow}
\usepackage{array}
\usepackage{subfigure}
\usepackage{morefloats}
\begin{document}

\title{An Information-Theoretic Approach to Optimize JWST Observations and Retrievals of Transiting Exoplanet Atmospheres}
\author{Alex R. Howe}
\affil{Department of Astronomy}
\affil{University of Michigan}
\affil{1085 S. University, Ann Arbor, MI 48109, USA}
\affil{arhowe@umich.edu}
\author{Adam Burrows}
\affil{Department of Astrophysical Sciences, Princeton University}
\affil{Peyton Hall, Princeton, NJ 08544, USA}
\affil{burrows@astro.princeton.edu}
\author{Drake Deming}
\affil{Department of Astronomy, University of Maryland}
\affil{College Park, MD 20742, USA}
\affil{ddeming@astro.umd.edu}
%\date{}

% % % % % % % % % % % % % % % % % % % % % % % % % % % % % % % % % % % % % % % % % % % % % % % % % % % % % % % % % % % % % % % % % % % % % %

\begin{abstract}

We provide an example of an analysis to explore the optimization of observations of transiting hot jupiters with {\it JWST} to characterize their atmospheres, based on a simple three-parameter forward model. We construct expansive forward model sets for eleven hot jupiters, ten of which are relatively well-characterized, exploring a range of parameters such as equilibrium temperature and metallicity, as well as considering host stars over a wide range in brightness. We compute posterior distributions of our model parameters for each planet with all of the available {\it JWST} spectroscopic modes and several programs of combined observations and compute their effectiveness using the metric of estimated mutual information per degree of freedom. From these simulations, clear trends emerge that provide guidelines for designing a {\it JWST} observing program. We demonstrate that these guidelines apply over a wide range of planet parameters and target brightnesses for our simple forward model.

\end{abstract}

\maketitle

% % % % % % % % % % % % % % % % % % % % % % % % % % % % % % % % % % % % % % % % % % % % % % % % % % % % % % % % % % % % % % % % % % % % % %

\section{Introduction}

The number of known extrasolar planets has grown dramatically in recent years, mostly due to the discoveries of the {\it Kepler} spacecraft \citep{2016ApJS..224...12C}. The majority of known exoplanets are transiting, making spectroscopic follow-up to determine their atmospheric physical properties a viable method of characterization for some. Because of selection biases and proximity to their host stars, the characterization of transiting planets favors ``hot jupiters''$-$giant planets in very short-period orbits$-$rather than the more typical ``mini-Neptunes'' and ``super-Earths'' discovered in large numbers by {\it Kepler}. The strong transit signatures and frequent, short transits of nearby ($<150$ pc) ``hot jupiters'' make them by far the easiest to characterize, and the most useful with which to test retrieval methods.

To characterize the atmospheres of transiting planets, one can measure the wavelength variation in transit depth, which provides information about atmospheric composition and scale height. It is also possible to measure secondary eclipse spectra, which provide information about emission in the infrared, and phase curves over an orbit, which shows longitudinal variation in temperature and chemistry. However, the most detailed spectra currently available cover only narrow bands such as those provided by the {\it Hubble Space Telescope (HST)}. These observations provide the general shape of the spectrum, especially at short wavelengths and frequently reveal the presence of water around 1.4 $\mu m$, but they provide little information about other species. Many other conclusions drawn from these data are often of dubious validity \citep{2014PNAS..11112601B}.

Moreover, observed primary transit spectrophotometry is often either featureless within observational uncertainties or shows only a rise in transit depth at blue wavelengths, indicating the presence of clouds or hazes that intercept the stellar light. A few major chemical features have been clearly detected$-$mainly the sodium and potassium lines at 0.59 $\mu m$ and 0.77 $\mu m$, respectively, along with the 1.4-$\mu m$ water feature. The rise in transit depth observed at short wavelengths is consistent with Mie scattering by micron-size haze particles; with very little mass, a trace species at a high altitude ($<0.01$ bar) can produce a rising transit spectrum at short wavelengths \citep{2016Natur.529...59S}.

The upcoming {\it James Webb Space Telescope (JWST)}, to be launched by the end of 2018, will offer an enormous advantage in spectroscopic observation of exoplanets over existing platforms such as {\it HST} and the {\it Spitzer Space Telescope}. It possesses a much larger collecting area (diameter 6.5 meters, as opposed to 2.4 meters for {\it HST}), and it will be able to observe spectra panchromatically from 0.6 $\mu m$ to 28.3 $\mu m$. It will also observe continuously, without interruption by Earth occultations. Unlike {\it HST}, which takes near-infrared spectra of exoplanets at a very low spectral resolution of $R\,(\Delta\lambda/\lambda)\sim 10-20$, all of {\it JWST's} spectroscopic modes will have at least R$\sim$100 and some as high as R$\sim$3000, with a possibility of R$\sim$2000 over the entire wavelength range. {\it JWST} has five scientific instruments, four of which will do spectroscopic observations. These are NIRCam \citep{2007SPIE.6693E..0GG}, NIRISS \citep{2012SPIE.8442E..2RD}, and NIRSpec \citep{2014SPIE.9143E..08B}, all of which will observe in the near-IR between 0.6 and 5.0 $\mu m$, plus MIRI, which will observe in the mid-IR from 5 to 28.3 $\mu m$ \citep{2015PASP..127..623K,2015PASP..127..646W}. The fifth instrument, FGS-Guider (0.8-5.0 $\mu m$) is a purely imaging instrument \citep{2012SPIE.8442E..2RD}. The angular resolution of the instruments is diffraction-limited in most configurations and ranges from 0.032 arcsec (NIRCam) to 0.11 arcsec (MIRI).

{\it JWST's} capabilities will be ideal for improving our knowledge of extrasolar planets by spectroscopic observation, equaling or surpassing all prior efforts in sensitivity and spectral resolution with much greater wavelength coverage. Judicious selection of targets and robust modeling efforts will be needed to most effectively use valuable telescope time. In this paper, we specifically focus on transit spectra, not secondary eclipse emission spectra or light curves, which will be covered in future works. Many studies of yields, targets, and retrievals of atmospheric properties of planets with JWST have been done \citep{2012A&A...538A..95M,2014A&A...563A.103S,2015MNRAS.448.2546B,2016ApJ...817...17G}, and it is likely many of the best hot jupiter targets have already been discovered by existing surveys such as SuperWASP \citep{2006PASP..118.1407P}, and HATNet \citep{2004PASP..116..266B}, although the upcoming {\it TESS} mission will turn greater attention to {\it JWST's} continuous viewing zone \citep{2009PASP..121..952D}.

In general, observations of planetary spectra may be inverted to obtain useful information about their atmospheres using various statistical retrieval techniques. These are compared with a model set of theoretical spectra computed over a range of atmosphere parameters. A best fit to the observations is found using a statistical algorithm, most commonly a Markov-Chain Monte Carlo algorithm (MCMC), but other algorithms may be used, including other Monte Carlo algorithms or non-Monte Carlo algorithms such as the optimal estimation method \citep{2000SAOPP...2.....R}. These algorithms provide a posterior distribution for the parameters of the atmosphere, as long as the correct answer is within the forward model set.

A number of such codes have been developed, each with a different emphasis. The BART code, produced by Harrington et al., is a highly modular code designed to compute spectra from first principles with thermal equilibrium molecular abundances. The CHIMERA secondary eclipse code by \citet{2013ApJ...775..137L} has a statistics package with multiple Monte Carlo methods and an optimal estimation method for optimization purposes. The NEMESIS code by \citet{2008JQSRT.109.1136I} is designed for a fast and accurate analysis of planetary spectra, particularly optimized for solar system objects, including additional parameters such as limb darkening, surface temperature, and detailed cloud models. The SCARLET code by \citet{2015arXiv150407655B} emphasizes a chemical and structural model by parameterizing 3-D atmosphere properties such as the eddy diffusion constant, heat-transport, and Bond albedo. Finally, the Tau-REx code by \citet{2015ApJ...802..107W} has a complex statistical package, which includes procedurally-generated priors.

Currently, about sixty exoplanets have been observed photometrically in primary transit in multiple wavebands. Most of these objects are hot jupiters, and most of them have been observed photometrically with the {\it Spitzer Space Telescope}. About twenty exoplanets have been spectroscopically characterized at low resolutions of $R\sim 10-20$, mostly with {\it HST} between 0.3 and 1.0 $\mu m$ with STIS and/or between 1.1 and 1.7 $\mu m$ with WFC3 \citep{2016Natur.529...59S}. Three objects have been studied in by far the greatest detail due to the brightness of their host stars and their ease of observation: HD 189733b \citep{2013MNRAS.432.2917P,2014ApJ...791...55M}, HD 209458b \citep{2013ApJ...774...95D}, and the mini-Neptune GJ 1214b \citep{2012ApJ...756..176H}.

{\it JWST} will be capable of many spectroscopic observing modes (a particular filter and other optical elements on a given instrument), giving many options to take observations. We may define an observing program as a particular combination of observing modes and integration times with which to observe a particular object or set of objects.\footnote{In this paper, we also use the phrase ``observing program'' to refer to our numbered observing programs in Section \ref{JWST} that are determined only by the observing modes used, while the integration time or number of transits per mode are specified independently.}

\begin{figure*}[htbp]
\includegraphics[width=0.99\textwidth]{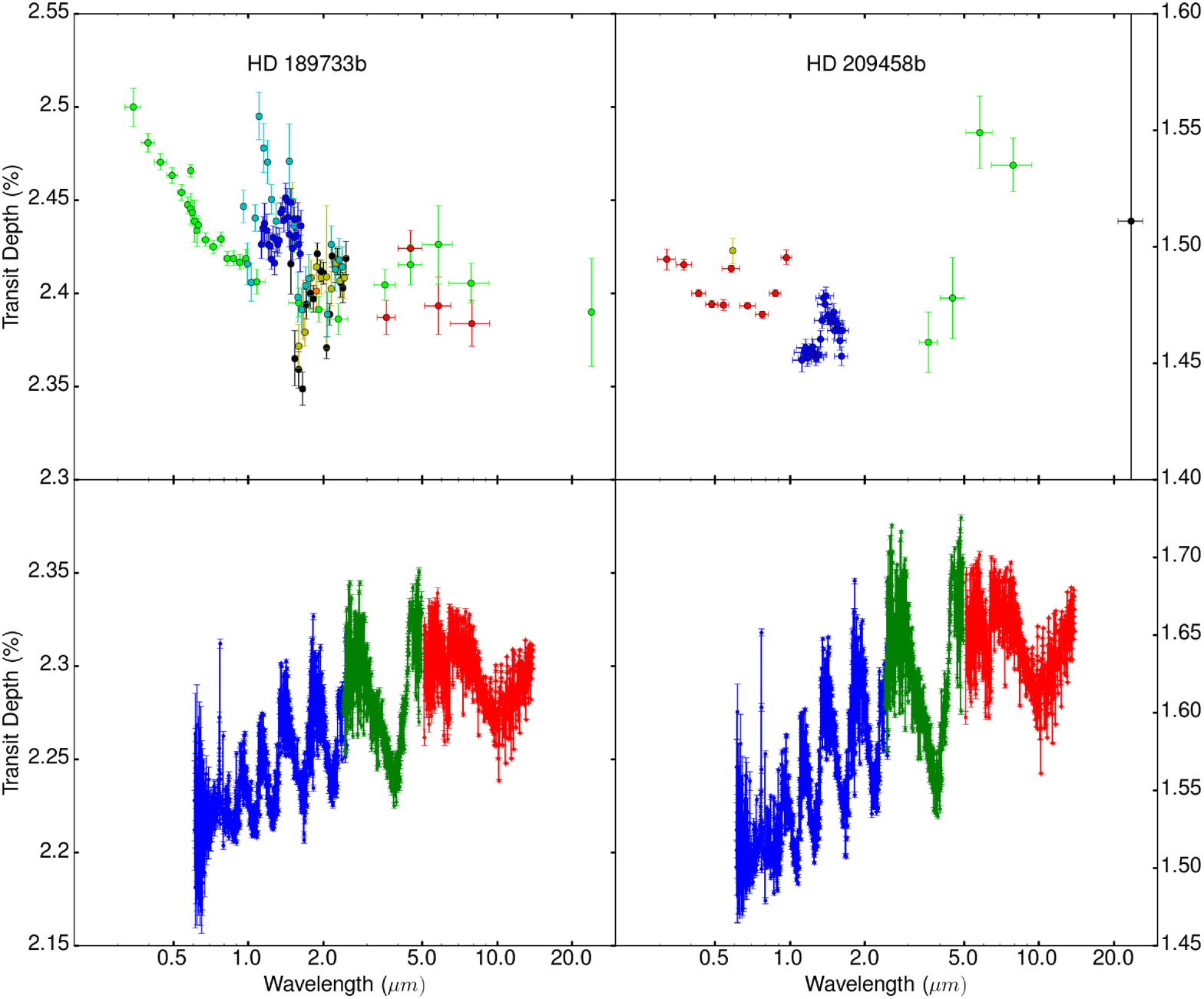}
\caption{Top: plots of the best published photometric and spectroscopic transit observations for HD 189733b and HD 209458b, the best-observed hot jupiters. Data for HD 189733b: \citet{2008Natur.452..329S}, black; \citet{2009ApJ...699..478D}, red; \citet{2009A&A...505..891S}, orange; \citet{2013ApJ...766....7W}, yellow; \citet{2013MNRAS.432.2917P}, green; \citet{2014ApJ...785...35D}, cyan; \citet{2014ApJ...791...55M}, blue. Data for HD 209458b: \citet{2002ApJ...568..377C}, black; \citet{2006ApJ...649.1043R}, red; \citet{2007Natur.448..143K}, yellow; \citet{2010MNRAS.409..963B}, green; \citet{2013ApJ...774...95D}, blue. Bottom: simulated transit spectra for the same objects with {\it JWST}: NIRISS in black and blue, NIRCam in green, MIRI in red. Anticipated error bars are included, but are very small and are not visible at all wavelengths. Since {\it JWST} has no sensitivity shortward of approximately 0.6 microns, we do not plot the theoretical spectra in this regime. Clearly, the current data are sparse and information-poor when compared with the potential of {\it JWST}.}
\label{plot189-209}
\end{figure*}

Plots of the best available transit observations in multiple wavelength ranges for the best-studied objects HD 189733b and HD 209458b are shown in the top panels of Figure \ref{plot189-209} to illustrate the quality of observations currently available. For comparison, simulated observations of the same objects with {\it JWST} are shown in the bottom panels. These simulated observations mostly have smaller error bars, more than an order of magnitude higher spectra resolution, and much more complete spectral coverage.

% Figure 1

With the much greater quality and quantity of spectroscopic data anticipated from {\it JWST} and the large number of options for observations, it is important to find the modalities that will make the best use of observing time. In this paper, we seek to explore these possibilities to determine quantitatively the observing programs that provide the strongest constraints on the model parameters with the least observing time and to develop more generally the methods to optimize observing programs of transiting planets with JWST. To simplify the problem and focus clearly on optimization, we employ a simple, three-parameter model and apply it to a wide variety of synthetic observations. We use this simple model to present a rigorous methodology for optimizing observing programs.

To evaluate our observing programs, we consider several statistical measures to find the best fit and to determine the amount of observing time needed to reach a certain level of precision. One important measure is the difference in Shannon entropy between the prior and posterior distributions in our retrievals, which is directly related to the size of the error bars in the Gaussian case and can be compared directly between different observations. However, a more complete measure of the difference between observing programs is the mutual information \citep{lindley1956,Chaloner95bayesianexperimental} between the prior and posterior distribution, which additionally accounts for the probability distribution of the observed data, thus marginalizing over the possible outcomes of the observing program. Therefore, we use this measure for our optimization of observing programs. The ideal prior would be a posterior from a previous observation, although we examine simple priors to study the effects of the choice of prior in the retrievals.

For this paper, we have developed a new transit spectra modeling and retrieval code, APOLLO, to perform such an analysis and developed techniques to determine the optimal strategies for observing transiting planets with {\it JWST} and a means of quantifying that optimization. We build a representative spectral model suite for eleven hot jupiters: HAT-P-1b, HAT-P-12b, HD 189733b, HD 209458b, Kepler-7b, WASP-6b, WASP-12b, WASP-17b, WASP-19b, WASP-39b, and WASP-43b. We then fit synthetic observations to them from a range of possible {\it JWST} observing programs using an MCMC algorithm to explore how to optimize an observing run.

The outline of this paper is as follows. We describe the information theory behind our analysis in Section \ref{info}. The observing capabilities of {\it JWST} for transits are explored in depth, and several observing programs are presented in Section \ref{JWST}. We describe the APOLLO code and the specific models we use in this study in Section \ref{our-code}. Our results and analysis with an information theoretic approach are provided in Section \ref{results}, and we list general conclusions in Section \ref{summary}.

% % % % % % % % % % % % % % % % % % % % % % % % % % % % % % % % % % % % % % % % % % % % % % % % % % % % % % % % % % % % % % % % % % % % % % %

\section{Information Theory}
\label{info}

A set of synthetic observations, $x$, can be used to derive a best fit, $\bar{\theta}$, and posterior distribution, $p(\theta|x)$, in the parameter space of the forward model set using an MCMC code. Based on this posterior distribution, we wish to know which observing program provides the greatest amount of information for the least possible observing time, that is, which set of observations allows us to estimate the model parameters to the greatest precision for the amount of telescope time required or available.

Before we compute rigorous statistical measures, we can examine the Jacobian, that is, the matrix of $\partial x_i/\partial \theta_j$, where $i$ and $j$ are the components of $x$ and $\theta$, respectively. These are the partial derivatives of the transit spectra with respect to the model parameters \citep{2013ApJ...775..137L}, and they provide a heuristic guideline for which wavelength bands are most useful to characterize planetary atmospheres. However, the value of examining Jacobians is likely to be more fully realized when looking at more complicated forward models for which they could highlight bands that serve as diagnostics for a single parameter out of a large set, such as the abundance of a particular subdominant molecular species. Alternatively, they could highlight diagnostics for cloud compositions, particle sizes, or other cloud properties that have been very difficult to determine with existing low-resolution spectrophotometric data. However, while we provide a demonstration of this technique, we do not apply this step in our analysis because we wish to consider all possible observing modes, and we do not need to narrow them down in this way.

Another possible metric for estimating the merit of an observing program at the intermediate stage is the Bayesian information criterion (BIC) \citep{2014MNRAS.444.3632H}, which is defined by:
\begin{equation}
BIC = -2\ln{\hat{L}} + k\ln{n},
\end{equation}
where $\hat{L}$ is the maximum of the likelihood function, $n$ is the number of data points, and $k$ is the number of model parameters. The BIC is a metric to determine goodness of fit to a model that corrects for overdetermination of the system. A larger likelihood results in a smaller BIC, and so we wish to minimize the BIC for a better fit. If the likelihood is a Gaussian, the BIC is equivalent to:
\begin{equation}
BIC = \chi^2 + k\ln{n},
\end{equation}
so it is equivalent to chi-squared minimization, except that models are penalized for overfitting. However, the BIC does not provide information on {\it how} good the fit is, only a fit with a specific optimization.

Ultimately, we wish to know which observations produce the narrowest constraints on the model parameters. However, this is more complicated than finding the smallest error bars. The error bars are not necessarily informative, for example, if we attempt to use them to characterize a multi-peaked distribution, and they are less helpful when we also wish to quantify the observing time needed to reach a certain level of precision. To do this, we must have a quantitative measure of the amount of information provided by an observation.

An observation $x$ with probability $p(x)$ carries an information content of $I = -\log_2 p(x)$ bits \citep{shannon48}. In other words, a minimum of $I$ bits are required to encode the value of $x$. We care about this because it effectively tells us how much information we know about $x$, where $x$ in this case may be the observed value of some property of the planet. The entropy, $H(X)$, of the complete probability distribution $p(x)$, per \citet{shannon48}, is the expectation value of $I$:
\begin{equation}
H(X) = -\sum_{x\in X} p(x)\log_2 p(x).
\end{equation}
This represents the average information needed to encode a value of $x$ from the domain $X$, which is the set of all possible observed values of $x$. For a continuous distribution, this becomes\footnote{In theory, the observations and model parameters in our study have continuous probability distributions, but in practice, we compute them with MCMC sampling, so our computations must use the discrete formulae.}
\begin{equation}
H(X) = -\int_{x\in X}p(x)\log_2 p(x)dx.
\end{equation}

One bit of information is a significant gain in a field where few definite conclusions can be drawn with confidence. For a Gaussian distribution, the entropy is $H(X) = \log_2(\sigma_X\sqrt{2\pi e})$, so a difference of one bit in entropy approximately corresponds to a factor of 2 in $\sigma$. For example, one of the prior distributions we employ in our study specifies an atmosphere metallicity of $0.5 \pm 0.5$ dex, or more specifically, a normal distribution with $\mu_{\log Z} = 0.5$ dex and $\sigma_{\log Z} = 0.5$ dex. If an observing program provides one bit of information, then we could use it to measure the metallicity of a planet's atmosphere to a precision of $\sigma_{\log Z} = 0.25$ dex. Similarly, a gain of $\log_2 10 = 3.32$ bits would add one significant (decimal) digit to the measurement.

For a given observation, the entropy of the posterior distribution of model parameters is $H(\Theta|x)$, which is evaluated over the domain of model parameters, $\Theta$. However, because we do not know {\it a priori} what the outcome of an observation will be, we must take the average entropy of the posterior over all possible observations to obtain an accurate measure of how informative an observation will be. We do this by marginalizing over the Bayesian evidence, $p(x)$ to obtain $H(\Theta|X)$:
\begin{align}
H(\Theta|X) &= \int_{x\in X} p(x)H(\Theta|x)dx \nonumber \\
            &= -\int_{x\in X} \int_{\theta\in \Theta} p(x)p(\theta|x)\log_2 p(\theta|x)d\theta dx,
\end{align}

The average amount of information obtained from a given observation is equal to the difference between the entropy of the prior and the posterior distribution, $H(\Theta) - H(\Theta|X)$, which is also known as the mutual information, $I(\Theta,X)$, a quantity that is familiar in the context of Bayesian experimental design \citep{lindley1956,Chaloner95bayesianexperimental,journals/ploscb/LiepeFKS13}. The formal definition of the mutual information is:
\begin{equation}
I(\Theta,X) = \int_\Theta \int_X p(x,\theta) \log_2\left(\frac{p(x,\theta)}{p(x)p(\theta)}\right) dxd\theta,
\end{equation}
where $p(x,\theta)$ is the joint probability distribution, defined as $p(x,\theta) = p(\theta|x)p(x) = p(x|\theta)p(\theta)$, from Bayes' Theorem:
\begin{equation}
p(\theta|x) = \frac{p(x|\theta)p(\theta)}{p(x)}.
\end{equation}
Here, $p(\theta)$ is the prior distribution of the model parameters, $\theta$, and $p(x|\theta)$ is the likelihood, determined by the noise levels for the observation. Again, $p(x)$ is the Bayesian evidence (which in practice is handled internally by an MCMC algorithm).

Note that the mutual information is also symmetric: $I(\Theta,X) = I(X,\Theta)$. If $X$ and $\Theta$ are completely uncorrelated, then $I(\Theta,X) = 0$. The mutual information can also be defined in terms of the Kullback-Leibler divergence \citep{Kullback51klDivergence}:
\begin{equation}
I(\Theta,X) = D_{KL}(p(x,y)||p(x)p(y)).
\end{equation}

To see that $I(\Theta,X) = H(\Theta) - H(\Theta|X)$, we use the definition of joint probability and rewrite the mutual information as:
\begin{align}
I(\Theta,X) &= \int_\Theta \int_X p(\theta|x)p(x) \log_2\left(\frac{p(\theta|x)p(x)}{p(x)p(\theta)}\right) dxd\theta \nonumber \\
            &= \int_\Theta \int_X p(\theta|x)p(x) \log_2\left(\frac{p(\theta|x)}{p(\theta)}\right) dxd\theta. \nonumber
\end{align}
Then, simplifying, we find that:
\begin{align}
I(\Theta,X) &= \int_\Theta \int_X p(\theta|x)p(x) \log_2 p(\theta|x) dxd\theta \nonumber \\
            &\ \ \ \ \ - \int_\Theta \int_X p(\theta|x)p(x) \log_2 p(\theta) dxd\theta \nonumber \\
            &= -H(\Theta|X) - \int_\Theta \log_2 p(\theta)\left(\int_X p(\theta|x)p(x)dx\right)d\theta \nonumber \\
%            &= -H(\Theta|X) - \int_\Theta p(\theta)\log_2 p(\theta) \left(\int_X p(x)dx\right)d\theta \nonumber \\
            &= -H(\Theta|X) - \int_\Theta p(\theta)\log_2 p(\theta)d\theta \nonumber \\
            &= -H(\Theta|X) + H(\Theta),
\end{align}
which completes the proof. Thus, it is the {\it difference} in entropy, which is dimensionless, that we care about rather than the entropy of the prior or posterior by itself, which, for a continuous probability density, is dependent on the units of the parameters.

Mutual information, much like the Shannon entropy, is derived from coding theory, and it finds significant use in the field of systems biology (e.g. \citealt{journals/ploscb/LiepeFKS13}), where it is similarly used to inform experimental design, and we propose that it is also appropriate for this optimization problem. However, we add the modification that it is the mutual information {\it per degree of freedom} that is the figure of merit. We justify this by noting that the amount of mutual information obtained for some of our model observations is large, sometimes exceeding 20 bits. We argue that such a value is reasonable because it is calculated for a three-parameter atmosphere model and it thus distributed over 3 degrees of freedom, so that 20 bits of total mutual information corresponds to 6.67 bits per degree of freedom for our model over and above the prior. This corresponds to error bars 100 times narrower than the width of the prior, which is reasonable for the very broad priors we use in this study. Most measurements will yield less mutual information than this, and there may be other systematic factors that prevent reaching this precision. For a more complex forward model with more parameters, the mutual information per degree of freedom is likely to be lower, and measuring the parameters to a precision of a few percent to a few tens of percent would be plausible based on these results.

However, the difficulty with the use of mutual information is that it is computationally intensive, especially for the dense data sets produced by {\it JWST}. To calculate it requires Monte Carlo integration over both the model parameters and the synthetic observations. Therefore, in order to simplify the calculation, we employ the approximation
\begin{align}
I(\Theta,X) &\approx H(\Theta) - H(\Theta|X) \approx H(\Theta) - H(\Theta|x) \nonumber \\
            &= - \int_\Theta p(\theta)\log_2 p(\theta)d(\theta) + \int_\Theta p(\theta|x) \log_2 p(\theta|x) d\theta.
\end{align}
This is just the difference in entropy between the prior and the posterior for a single simulated observation rather than the average, computed without marginalizing over the evidence. This is also the metric used by \citet{2012ApJ...749...93L} in their analysis. Since the observation contains many (hundreds or more) data points, the effect of outliers on the conditional entropy estimate should mostly damp out. We estimate the accuracy of this approximation in Section \ref{error} by computing a low-resolution Monte Carlo integral of the full mutual information for comparison. We do this by computing many posterior distributions from different random seeds with a Gaussian distribution and averaging the results. Since the formula using a single posterior distribution is an approximation to the correct value, we use it as an estimate, which we define as the {\it posterior entropy method}.

We also have a second method to estimate the mutual information. For this, we assume that the posterior is a multivariate normal distribution. In this case, the mutual information is:
\begin{equation}
I(\Theta,X)\approx \frac{1}{2}\log_2\left(\frac{|A+C|}{|C|}\right),
\label{covar}
\end{equation}
where $A$ is the covariance matrix of the prior, and $C$ is the covariance matrix of the posterior. This is a generalization of the single-variable case:
\begin{equation}
I(\Theta,X)\approx \frac{1}{2}\log_2(1+\tau^2/\sigma^2),
\end{equation}
where $\tau$ is the standard deviation of the prior, and $\sigma$ is the standard deviation of the posterior \citep{shannon48,lindley1956}. If the probability distributions are Gaussians, Equation \ref{covar} computes the mutual information exactly, and it should be a good approximation to mutual information for non-Gaussian distributions. We define this formula as the {\it covariance matrix method} for estimating the mutual information in the general case. This formula proceeds directly from the magnitude of the error bars, which is ultimately a figure of interest. In order to determine how accurate this method is for non-Gaussian distributions, we again average the results of many calculations from different random seeds.

%We have one other method that we can use to help determine the optimal observing program. When we examine the difference in mutual information for particular observing programs, we can compare this difference across other parameters. For example, when we examine the difference between using a flat and a Gaussian prior, we can take the difference in the information gain between the two models and compute this difference for every observing mode and integration time and average them to determine the expected advantage of one prior over the other. This is essentially an alternative averaging method, and for appropriate parameters, the difference in entropy can be very consistent across observing modes. For example, when comparing a flat and a Gaussian prior across all of our observing programs for HAT-P-1b, an observing program modeled with a Gaussian prior has an average of 0.04 bits more entropy than the same observing program modeled with a flat prior, with a standard deviation of 0.18 bit (see Section \ref{error}).

% % % % % % % % % % % % % % % % % % % % % % % % % % % % % % % % % % % % % % % % % % % % % % % % % % % % % % % % % % % % % % % % % % % % % % %

\section{JWST Capabilities}
\label{JWST}

\begin{figure*}[htp]
\includegraphics[width=0.99\textwidth]{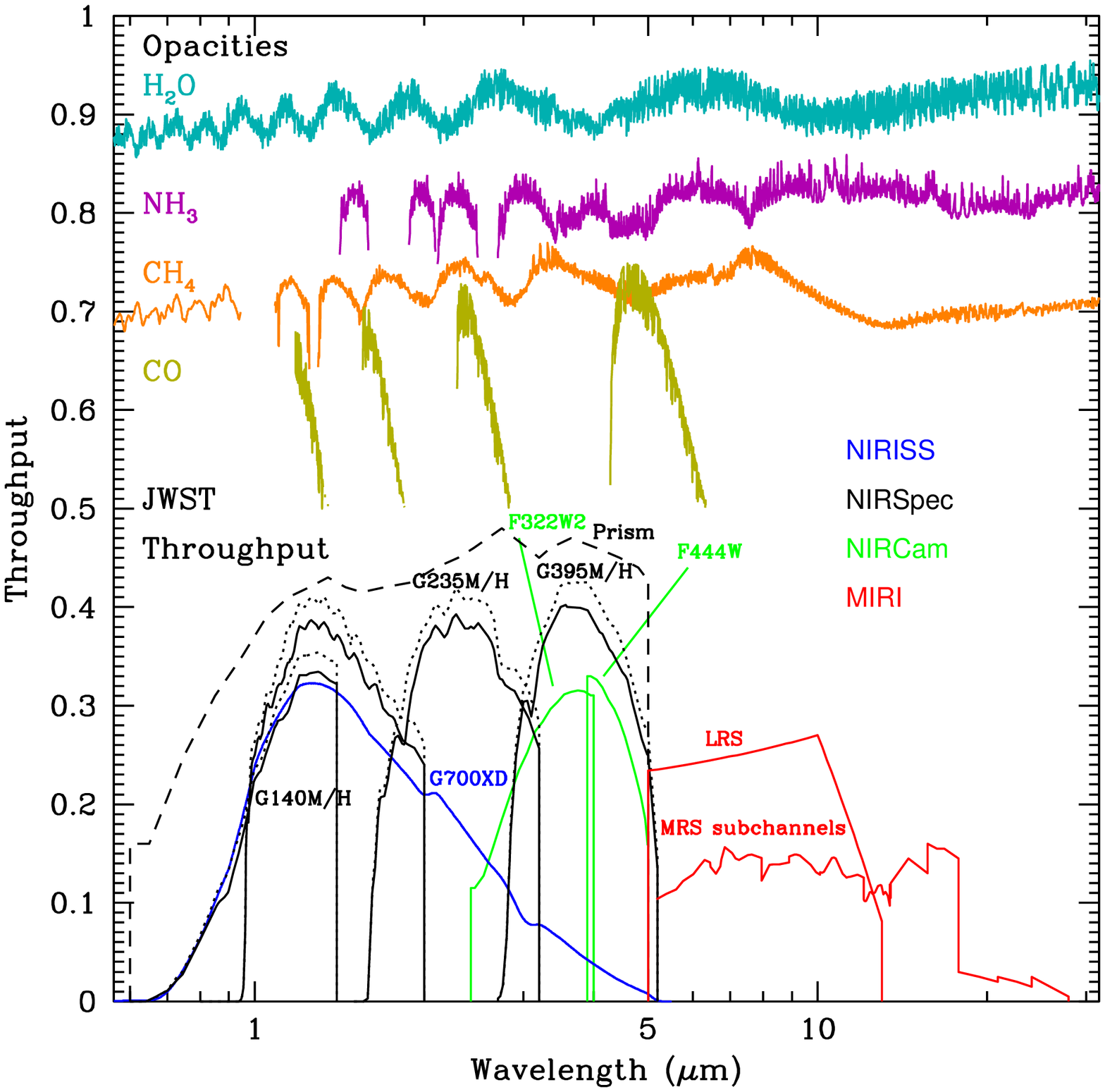}
\caption{Throughput functions of all JWST spectroscopic observing modes versus wavelength in microns plotted with common molecular opacities (scaled and offset). For NIRSpec, high-resolution modes are shown with solid lines, while medium-resolution modes are shown with dotted lines. The G140M and G140H grisms are each used in two observing modes corresponding to two filters. MIRI LRS corresponds to both the Slit and Slitless modes. MIRI MRS is divided into twelve subchannels, which can be imaged in three visits.}
\label{JWSTPlot}
\end{figure*}

JWST will have many observing modes for photometry, spectroscopy, and imaging. Of these, the spectroscopy modes are the most relevant for planetary atmosphere characterization via transit spectra since spectroscopy is needed to identify molecular features. As currently planned, JWST will have fifteen spectroscopic modes covering ten wavelength ranges between its four instruments. These modes are listed in Table \ref{JWSTModes}, and their throughput functions are plotted alongside major molecular opacity spectra in Figure \ref{JWSTPlot}. The high-resolution NIRSpec modes are shown as solid lines, while the medium-resolution modes are shown as dotted lines. Note that NIRSpec's G140M and G140H grisms can operate with two different filters, and that MIRI MRS is divided into twelve subchannels, which require three visits to take a complete spectrum. There are also two modes for MIRI LRS: the Slit and Slitless modes. However, these cover the same wavelength range at similar resolution, so they are not distinguished here.

% Table 1
\begin{table*}[htbp]
\caption{JWST Spectroscopic Modes}
\begin{center}
\begin{tabular}{l|l|l|l|l}
\hline
Instrument & Mode         & Wavelength Range ($\mu m$) & Resolution & Saturation Limit \rule{0pt}{2.6ex} \rule[-1.2ex]{0pt}{0pt} \\ [+2pt]
\hline
NIRCam     & F322W2       & 2.4-4.0   & R$\approx$1700      & K$\approx$4.6 \rule{0pt}{2.6ex} \\ [+2pt]
NIRCam     & F444W        & 3.9-5.0   & R$\approx$1700      & K$\approx$3.8 \rule[-1.2ex]{0pt}{0pt} \\ [+2pt]
\hline
NIRISS     & GR700XD      & 0.6-2.8$^1$   & R$\approx$430-1350  & J$\approx$6.35$^1$ \rule{0pt}{2.6ex} \rule[-1.2ex]{0pt}{0pt} \\ [+2pt]
\hline
NIRSpec    & F070LP+G140M & 0.7-1.2   & R$\approx$500-1300  & J$\approx$8.5 \rule{0pt}{2.6ex} \\ [+2pt]
NIRSpec    & F070LP+G140H & 0.7-1.2   & R$\approx$1500-3500 & J$\approx$7.5  \\ [+2pt]
NIRSpec    & F100LP+G140M & 1.0-1.8   & R$\approx$500-1300  & J$\approx$8.5  \\ [+2pt]
NIRSpec    & F100LP+G140H & 1.0-1.8   & R$\approx$1500-3500 & J$\approx$7.5  \\ [+2pt]
NIRSpec    & F170LP+G235M & 1.7-3.0   & R$\approx$700-1300  & J$\approx$8.0  \\ [+2pt]
NIRSpec    & F170LP+G235H & 1.7-3.0   & R$\approx$2000-3500 & J$\approx$7.0  \\ [+2pt]
NIRSpec    & F290LP+G395M & 2.9-5.0   & R$\approx$700-1300  & J$\approx$7.0  \\ [+2pt]
NIRSpec    & F290LP+G395H & 2.9-5.0   & R$\approx$2000-3500 & J$\approx$6.0  \\ [+2pt]
NIRSpec    & CLEAR+PRISM  & 0.6-5.0   & R$\approx$30-100    & J$\approx$11.0 \rule[-1.2ex]{0pt}{0pt} \\ [+2pt]
\hline
MIRI       & LRS, Slit    & 5.0-13.0  & R$\approx$100       & K$\approx$5.7 \rule{0pt}{2.6ex} \\ [+2pt]
MIRI       & LRS, Slitless & 5.0-13.0 & R$\approx$100       & K$\approx$5.7    \\ [+2pt]
MIRI       & MRS1         & 5.0-7.7   & R$\approx$800-2400  & K$\approx$4    \\ [+2pt]
MIRI       & MRS2         & 7.7-11.9  & R$\approx$800-2400  & K$\approx$4    \\ [+2pt]
MIRI       & MRS3         & 11.9-18.4 & R$\approx$800-2400  & K$\approx$4    \\ [+2pt]
MIRI       & MRS4         & 18.4-28.3 & R$\approx$800-2400  & K$\approx$4 \rule[-1.2ex]{0pt}{0pt} \\ [+2pt]
\hline
\end{tabular}
\end{center}
\caption{$^1$Saturation limit $J=8.05$ for both orders, covering 0.6-2.8 microns; $J=6.35$ for first order, covering 0.85-2.8 microns.}
\label{JWSTModes}
\end{table*}

\begin{figure*}[htp]
\includegraphics[width=0.99\textwidth]{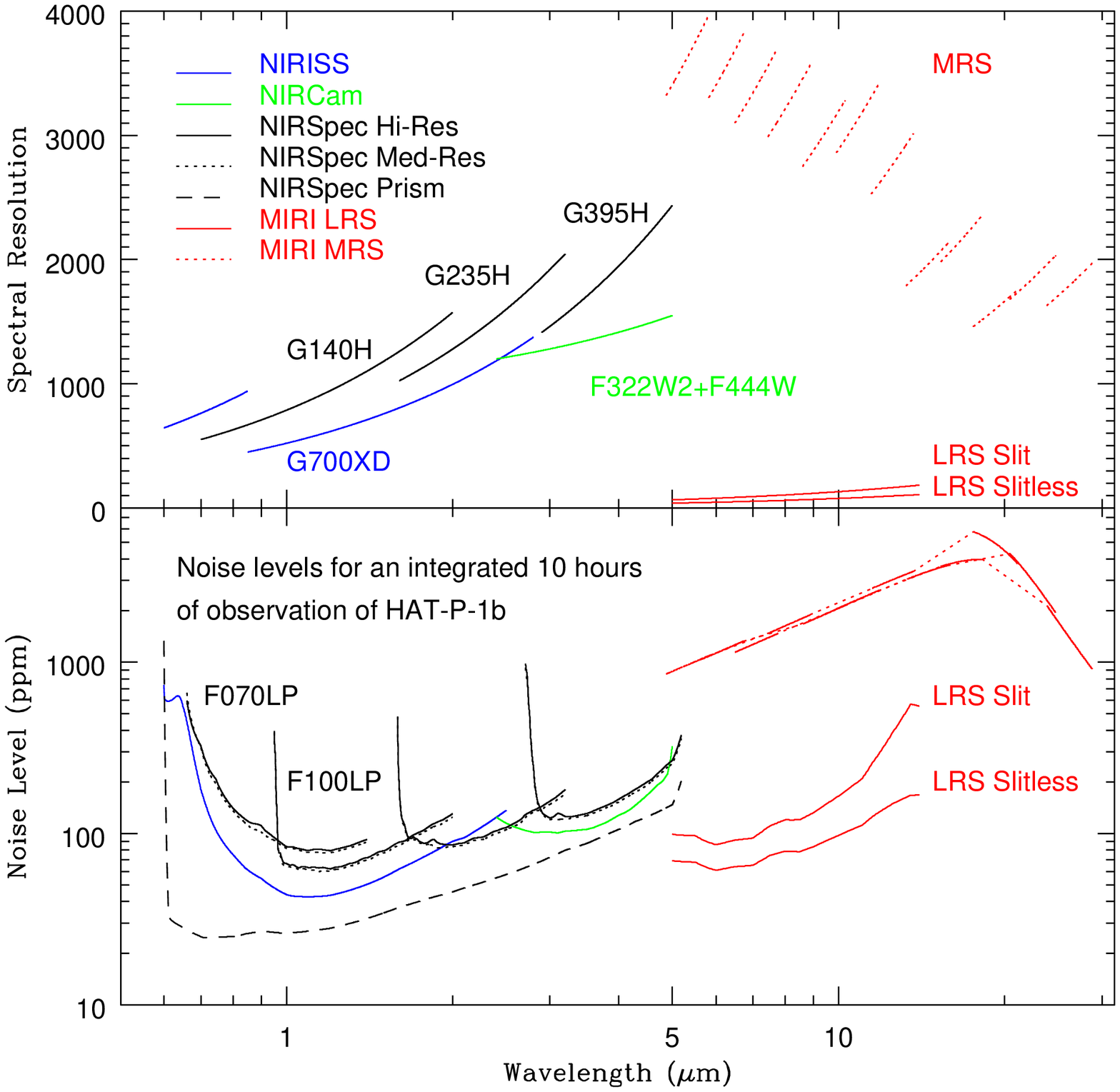}
\caption{Top: Spectral resolution versus wavelength in microns of all {\it JWST} spectroscopic observing modes. Rather than individual filters, the resolution curves correspond to grisms or other dispersion elements on the instruments. Thus, NIRSpec has three resolution curves, NIRCam has one, and MIRI MRS has four piecewise curves, corresponding to its four IFUs. Bottom: Noise levels in parts per million versus wavelength in microns for synthetic observations of a 10-hour integration on HAT-P-1 in all {\it JWST} spectroscopic observing modes. Note that NIRSpec Prism will not observe the same objects as the other modes, but these noise levels are still representative if the integrated light collection is the same. The MIRI Slit and Slitless modes are distinguished in this figure because they have different sky background levels.}
\label{JWSTNoise}
\end{figure*}

The spectral resolution of these modes is plotted in the top panel of Figure \ref{JWSTNoise}. In this case, it is not the curves corresponding to the individual filters that are shown, but rather those for the grisms and other dispersive elements. Elements that can be used with more than one filter, and thus more than one observing mode, are represented by a single curve. There is one such curve for NIRCam, three for NIRSpec, and four for MIRI MRS, corresponding to its four IFUs. The noise levels for representative synthetic observations (in this case, for a 10-hour integration on HAT-P-1) are shown in the bottom panel of Figure \ref{JWSTNoise} in terms of the uncertainty in the observed transit depth in parts per million. Note that the NIRSpec Prism will observe fainter stars than HAT-P-1, but we have estimated the noise levels using a linear model, so they should be similar to those for longer integrations of fainter stars. The MIRI LRS Slit and Slitless modes are distinguished here, with the Slitless mode having lower noise levels because of its higher throughput. For much fainter objects, the noise level for the Slitless mode will be higher because the background from the sky will be relatively higher without a slit.

NIRISS has one spectroscopic observing mode at medium resolution, the SOSS mode using the G700XD grating \citep{2012SPIE.8442E..2RD}. NIRCam has two modes at longer wavelengths, using the F322W2 and F444W filters and grisms, and the two instruments together efficiently cover the wavelength range from 0.6 to 5.0 $\mu m$ in three visits \citep{2007SPIE.6693E..0GG}. NIRISS and NIRCam are the best suited instruments for observing bright objects, with saturation limits of $J\approx 6.35$ and $K\approx 4.6$, respectively, making them capable of observing HD 189733b and HD 209458b, particularly with the four-amplifier, striping readout mode for NIRCam \citep{2014PASP..126.1134B}. This will be important for future {\it TESS} targets, since {\it TESS} will focus on bright stars. NIRCam also occupies an intermediate resolution between the NIRSpec/NIRISS medium-resolution modes and the NIRSpec high-resolution modes.

NIRSpec itself has three medium-resolution ($R=500-1300$) and three high-resolution ($R=1500-3500$) gratings, with each set able to cover the same wavelength range of 0.6 to 5.0 $\mu m$ in four visits \citep{2014SPIE.9143E..08B}. The high-resolution modes are more suited for observing objects of intermediate brightness, with saturation limits of $J\sim 7-8$. Thus, NIRSpec would not be able to observe HD 189733b and HD 209458b, but would be able to observe the other objects in our study. NIRSpec also has a Prism mode that covers the same wavelength range in one visit, but at a much lower resolution of $R\sim 100$ and a low saturation limit around $J\sim 11$.

MIRI has two spectrographs$-$one at low resolution from 5 to 14 $\mu m$ (LRS) \citep{2015PASP..127..623K} and one at medium-to-high resolution from 5 to 28 $\mu m$ (MRS) \citep{2015PASP..127..646W}. Both MIRI spectrographs are suitable for bright objects, with saturation limits of $K\approx 4-6$. However, MIRI MRS is composed of four integral field units (IFUs) of different wavelength ranges, each of which is split into three sub-channels, which must be imaged individually. Thus, MIRI requires three visits for complete wavelength coverage. MIRI MRS also faces unique challenges, including lower-than-expected efficiency and more complicated systematics than the other modes \citep{2015PASP..127..686G}.

All of the other spectroscopic observing modes for the four instruments require one visit each. As designed, JWST does not have parallel observation capability. The telescope hardware will allow it, and studies are being done to implement this capability \citep{2014PASP..126.1134B}, but this does not allow multiple observations of the same target at the same time. One proposal to implement truly simultaneous observations of the same target, due to \citet{2017PASP..129a5001S}, would involve adding an additional, short-wavelength spectroscopy channel to NIRCam.

The different observing modes are useful for different types of observations. For example, the only instruments that can take high resolution (R$\gtrsim$1500) spectra are NIRSpec in its high-resolution modes and MIRI MRS. The rest of the modes are relatively similar to each other in functionality, except for the NIRSpec Prism and MIRI LRS, both of which are low resolution (R$\sim$100). The NIRSpec Prism mode is also uniquely suited for faint targets ($J>11$), while brighter targets will saturate it.

It is important to select observing modes for JWST to make efficient use of observing time. Each observing mode (except MIRI MRS) requires one visit, consisting of one or more transits (in the case of transit spectroscopy). Therefore, we wish to cover as large a wavelength range as possible with as few visits as possible. Four particular observing modes together, NIRISS GR700XD, NIRCam F322W2, NIRCam F444W, and MIRI LRS, are the fewest modes that can achieve complete coverage of the wavelength range of interest (0.6 to 14 $\mu m$) for planets orbiting bright stars. The additional range covered by MIRI MRS (the 14.0 to 28.3 $\mu m$ range not covered by LRS) has not been frequently considered, but is likely to yield useful diagnostics to trace molecular species.

\begin{figure*}[htp]
\includegraphics[width=0.99\textwidth]{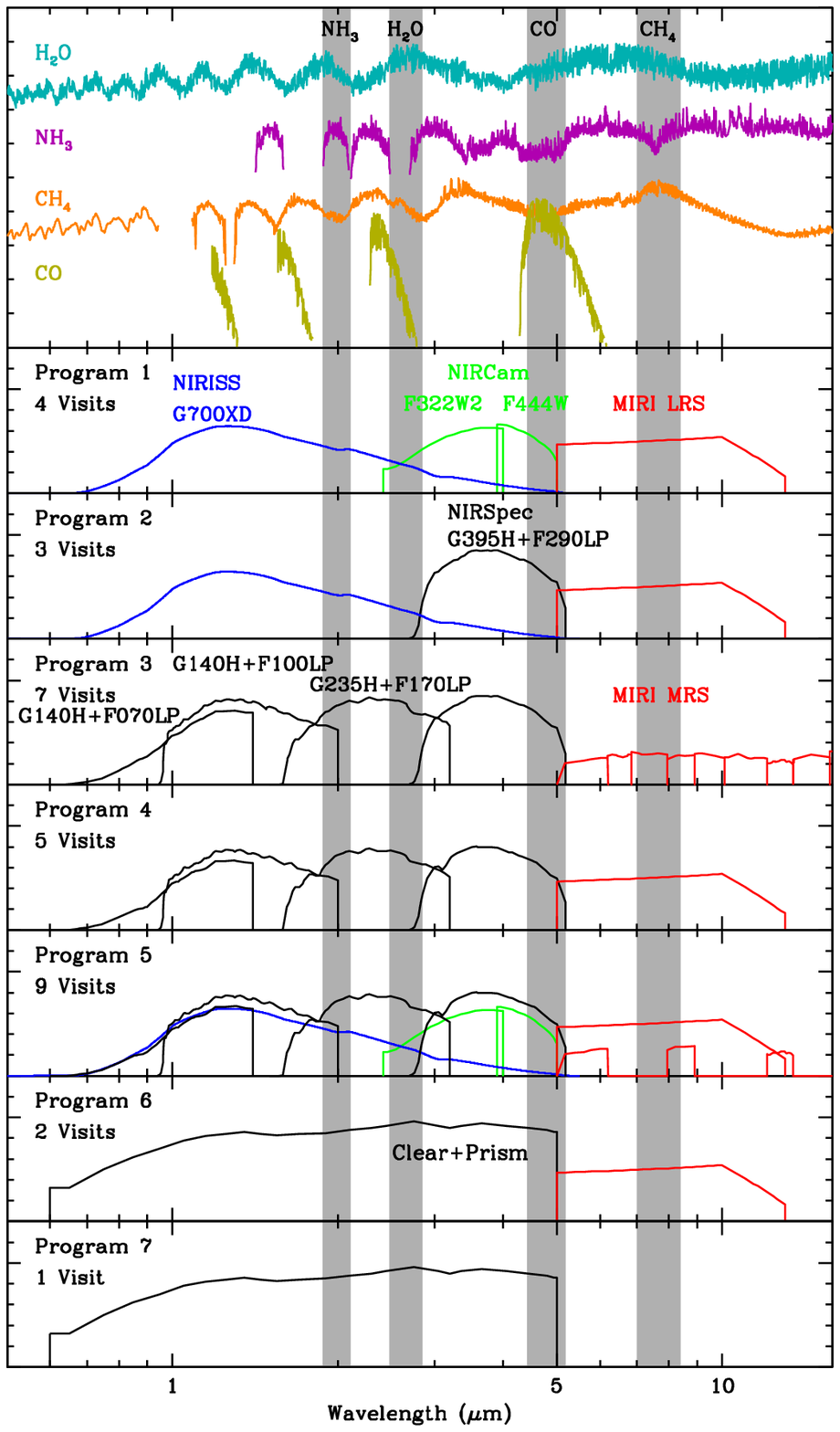}
\caption{Plots of throughput functions versus wavelength in microns of observing modes for seven model observing programs of interest to this paper, plotted with common molecular opacities. Several important molecular features are shaded. All of the programs are designed to cover most or (usually) all of the wavelength range from 0.6 to 14 $\mu m$. Program 1 covers the entire range at medium spectral resolution, while Program 3 does the same at high resolution. Programs 2 and 4 are designed to do nearly as much as Programs 1 and 3, respectively, with significantly less observing time, while Program 5 is designed to be as thorough as possible. Programs 6 and 7 are designed specifically for faint objects ($J>11$).}
\label{ComboPlot}
\end{figure*}

We summarize seven model observing programs in Table \ref{FilterCombos} and plot their throughput functions with major molecular opacities in Figure \ref{ComboPlot}. In all cases, MIRI LRS refers to the Slitless model, which we find to be more informative for our simple, three-parameter model, based on our analysis in Section \ref{LRS}. Program 1 images the entire spectrum from 0.6 to 14 $\mu m$ at low-to-medium resolution in four visits (using NIRISS GR700XD, NIRCam F322W2, NIRCam F444W, and MIRI LRS). This appears to be a well-optimized combination because of the relatively small number of observing modes, very little overlap between the modes, and high saturation limits for all four modes. Program 2 comprises three visits using NIRISS, NIRSpec G395M, and MIRI LRS. This program is designed to do nearly as much as Program 1 with one quarter less observing time, maintaining the full wavelength coverage, but with a lower saturation limit.

% Table 2
\begin{table*}[htb]
\caption{Representative Observing Programs}
\begin{center}
\begin{tabular}{l|l|r|l}
\hline
Instrument & Mode         & Visits & Notes \rule{0pt}{2.6ex} \rule[-1.2ex]{0pt}{0pt} \\
\hline
%\noalign{\vskip 2pt}
NIRISS     & GR700XD      & 4      & Program 1                   \rule{0pt}{2.6ex} \\ [+2pt]
NIRCam     & F322W2       &        & Mostly med-res over full range of interest.   \\ [+2pt]
NIRCam     & F444W        &        &                                               \\ [+2pt]
MIRI       & LRS          &        &                       \rule[-1.2ex]{0pt}{0pt} \\ [+2pt]
\hline
NIRISS     & GR700XD      & 3      & Program 2                   \rule{0pt}{2.6ex} \\ [+2pt]
NIRSpec    & F290LP+G395H &        & Fast, nearly-complete trade-off program       \\ [+2pt]
MIRI       & LRS          &        & for bright targets.   \rule[-1.2ex]{0pt}{0pt} \\ [+2pt]
\hline
NIRSpec    & F070LP+G140H & 7      & Program 3                   \rule{0pt}{2.6ex} \\ [+2pt]
NIRSpec    & F100LP+G140H &        & Fully high-res spectrum of interest.          \\ [+2pt]
NIRSpec    & F170LP+G235H &        & Best for medium-brightness targets.           \\ [+2pt]
NIRSpec    & F290LP+G395H &        &                                               \\ [+2pt]
MIRI       & MRSA         &        &                                               \\ [+2pt]
MIRI       & MRSB         &        &                                               \\ [+2pt]
MIRI       & MRSC         &        &                       \rule[-1.2ex]{0pt}{0pt} \\ [+2pt]
\hline
NIRSpec    & F070LP+G140H & 5      & Program 4                   \rule{0pt}{2.6ex} \\ [+2pt]
NIRSpec    & F100LP+G140H &        & Mostly high-res trade-off program.            \\ [+2pt]
NIRSpec    & F170LP+G235H &        &                                               \\ [+2pt]
NIRSpec    & F290LP+G395H &        &                                               \\ [+2pt]
MIRI       & LRS          &        &                       \rule[-1.2ex]{0pt}{0pt} \\ [+2pt]
\hline
NIRISS     & GR700XD      & 9      & Program 5 \rule{0pt}{2.6ex} \\ [+2pt]
NIRCam     & F322W2       &        & Maximum coverage for testing purposes.        \\ [+2pt]
NIRCam     & F444W        &        &                                               \\ [+2pt]
NIRSpec    & F070LP+G140H &        &                                               \\ [+2pt]
NIRSpec    & F100LP+G140H &        &                                               \\ [+2pt]
NIRSpec    & F170LP+G235H &        &                                               \\ [+2pt]
NIRSpec    & F290LP+G395H &        &                                               \\ [+2pt]
MIRI       & LRS          &        &                                               \\ [+2pt]
MIRI       & MRSA         &        &                       \rule[-1.2ex]{0pt}{0pt} \\ [+2pt]
\hline
NIRSpec    & CLEAR+PRISM  & 2      & Program 6 \rule{0pt}{2.6ex} \\ [+2pt]
MIRI       & LRS          &        & Fast, low-res program over range of interest. \rule[-1.2ex]{0pt}{0pt} \\ [+2pt]
\hline
NIRSpec    & CLEAR+PRISM  & 1      & Program 7 \rule{0pt}{2.6ex} \rule[-1.2ex]{0pt}{0pt} \\ [+2pt]
\hline
\end{tabular}
\end{center}
\label{FilterCombos}
\end{table*}

Program 3 covers the entire range from 0.6 to 28.3 $\mu m$ at high resolution. This requires seven visits for full coverage: four with NIRSpec and three with MIRI MRS. NIRISS and NIRCam appear to be more versatile than NIRSpec for transit observations because they are slitless, have finer spatial resolution, and have brighter flux limits than NIRSpec, but NIRSpec has much higher spectral resolution \citep{2016ApJ...817...17G}. Program 4 is again designed to do nearly as much as Program 3 with less observing time by substituting MIRI LRS for MRS and reducing the number of visits to five (four with NIRSpec and one with MIRI).

\citet{2016PASP..128i4401S} propose a {\it JWST} Early Release Science (ERS) Cycle 1 program to observe one transiting planet in all of the ``recommended'' observing modes to provide data to the community to improve future observation programs. This is important in part because no transit-spectroscopy-specific error budget for {\it JWST} has been specified, and detailed observations are needed to quantify the errors. This proposed program would use seven visits totaling roughly 60 hours for the three NIR instruments (using only one set of the NIRSpec modes without the prism), and two for MIRI (with only one visit with MRS for testing purposes), with an option to add two additional visits to cover the remaining MIRI subchannels. The best candidate for this observing program is WASP-62b. We specify this program, without the additional two visits, as Program 5.

For faint targets ($J>11$), the two low-resolution observing modes are of interest: NIRSpec Prism and MIRI LRS. We present the two of these modes together as Program 6, and NIRSpec Prism by itself as Program 7. This allows us to assess the expected productivity of different options for a potential campaign of short, low-resolution visits to many faint objects to gather a large dataset.

% % % % % % % % % % % % % % % % % % % % % % % % % % % % % % % % % % % % % % % % % % % % % % % % % % % % % % % % % % % % % % % % % % % % % %

\section{Outline of Our Retrieval Code}
\label{our-code}

\subsection{The APOLLO Transit Code}

APOLLO is a primary transit spectrum modeling and retrieval code with capabilities to expand to secondary eclipses and light curves. It particularly emphasizes a modular design that allows flexibility in the parameterization of the planetary atmosphere models, as well as the careful analysis of the relative information content of observations. The code generates a model set of transit spectra over a range of atmospheric parameters. With the modular design, the parameterization can potentially range from the very basic three-parameter model we use in this paper (consisting of a uniform metallicity, an isothermal temperature, and an opaque cloud deck), to a model with detailed temperature, composition, and cloud profiles of several types, and with cloud optical properties computed by Mie scattering. Our three-parameter model is a reasonable first-order approximation to real hot jupiter spectra because the assumptions of uniform metallicity and isothermal temperature are not far off from simulated atmosphere profiles in the upper layers that are probed by primary transit spectroscopy.

The transit spectrum computation is performed with an expanded version of our transit code from \citet{2012ApJ...756..176H}. Molecular opacities are derived from \citet{2007ApJS..168..140S} and pre-computed line-by-line on a grid of temperatures and pressures. The code can accept multiple treatments of the atmosphere composition, including a uniform composition with a multiple of solar metallicity and a composition with specified abundances of major molecular species. While the code allows the implementation of more complex profiles, for the purposes of this paper, we use an isothermal temperature profile.

Clouds are the most complex and least-understood part of the atmosphere model, and here, too, we allow for a number of parameterizations. The simplest parameterization is to place a completely opaque cloud deck at a given pressure level, as in this paper. A more complicated four-parameter model is also possible, with top and bottom cloud pressure levels, a modal particle size, and a uniform particle density, along with an extinction spectrum for the cloud material based on Mie scattering.

\begin{figure}[htbp]
\includegraphics[width=0.49\textwidth]{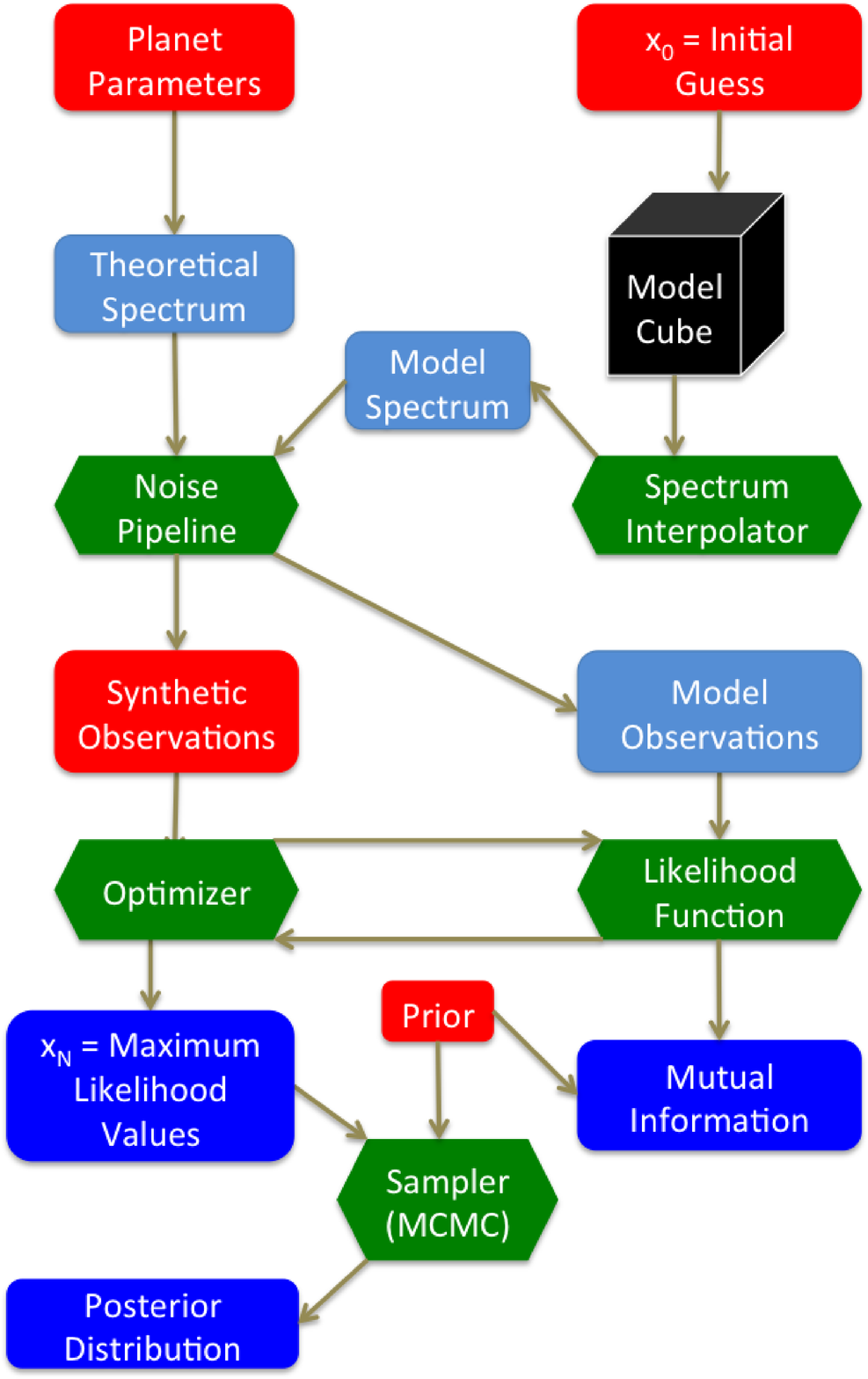}
\caption{Program flow of our transit code and statistical analysis.}
\label{Flowchart}
\end{figure}

The program flow of our code is shown in Figure \ref{Flowchart}.

% % % % % UPDATE

% % % % % UPDATE

\subsection{The Forward Model Set}

%We seek paths to develop the methods to optimize observing programs of transiting planets with JWST. Because we wish to explore techniques of optimization in depth rather than complicated or detailed forward models

We develop a simple, three-parameter model with large forward model sets for each object to which we can fit as wide as possible a variety of synthetic observations. Our three-parameter model includes a uniform temperature, a uniform metallicity, and an opaque cloud deck with a ceiling at a given pressure level. We compute a broad model set for each object based on its mass and radius and the properties of the host star to ensure that our retrieval algorithm can successfully retrieve the model's parameters from a wide range of possibilities. Our model sets include six metallicity points spaced logarithmically from -1.0 to 1.5 dex compared with solar abundances, 27 cloud top pressure points spaced logarithmically from $10^{-4}$ bar to $10^{2.5}$ bar, and a varying number of temperature points spaced linearly by either 50 K or 100 K for several hundred K around the equilibrium temperature. (The equilibrium temperature is the ``effective'' blackbody temperature at the planet's orbital distance assuming zero Bond albedo and complete heat redistribution.)

\begin{figure*}[htp]
\includegraphics[width=0.95\textwidth]{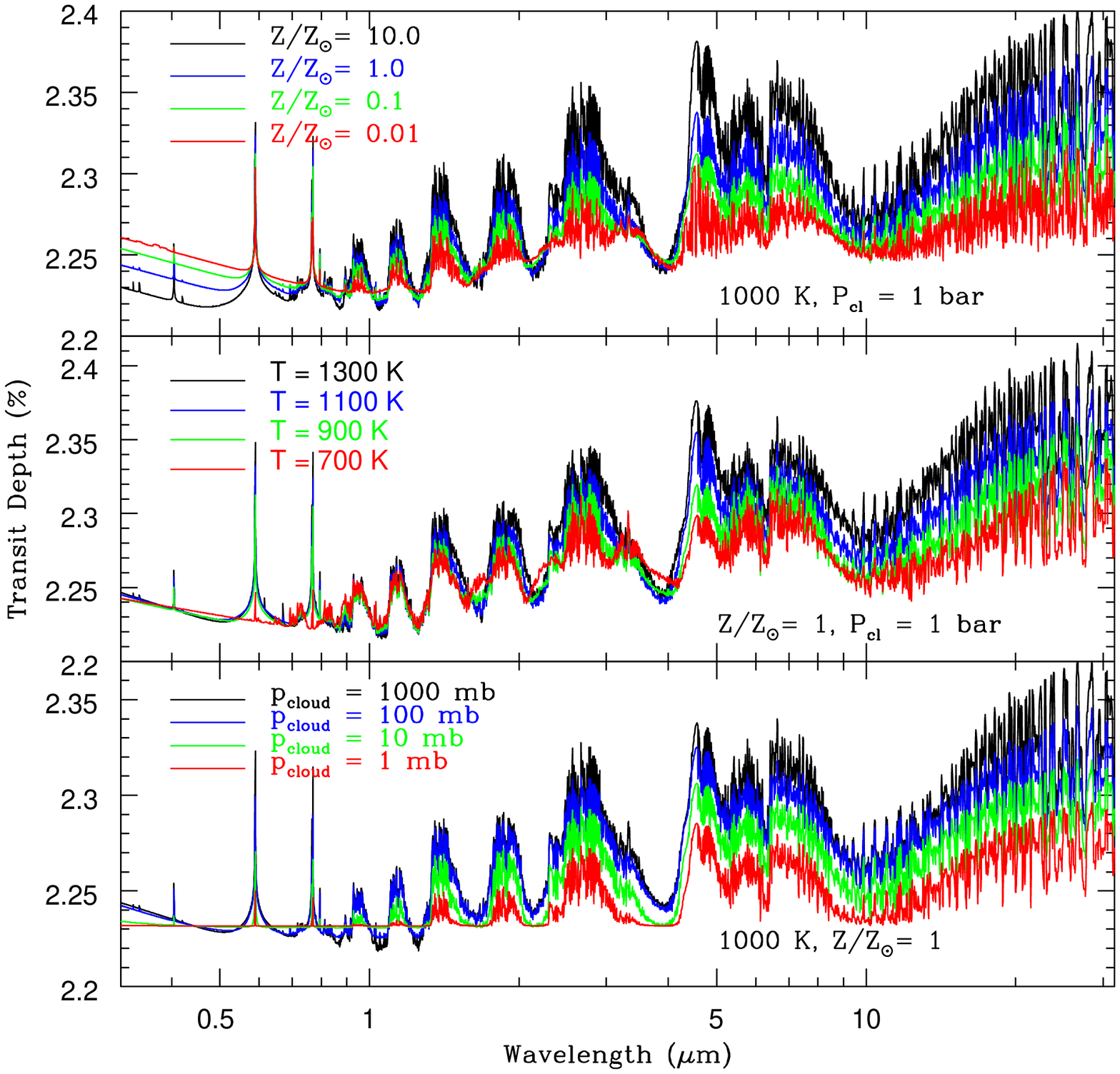}
\caption{Comparison of transit spectra in percent transit depth versus wavelength in microns for exoplanets with the physical parameters of HD 189733b and varying metallicities from -2.0 dex to 1.0 dex (top), isothermal temperatures from 700 K to 1300 K (middle), and opaque cloud decks with varying pressure levels from 1000 mb to 1 mb (bottom).}
\label{CompareZTP}
\end{figure*}

To demonstrate the effect of the input parameters on the spectra, we plot in Figure \ref{CompareZTP} a range of forward model transit spectra using the physical parameters of HD 189733b. The top panel shows the variation with planet metallicity. A planet with a higher metallicity will have larger transit depth spectral features with nearly the same overall shape despite the decrease in scale height, since the chemistry does not change greatly with metallicity. The largest change is the suppression of molecular features blueward of 1 $\mu m$ at low metallicities, where Rayleigh scattering becomes dominant. A global offset occurs due to the choice of normalization wavelength, near 1 $\mu m$, which results in the Rayleigh tail being pushed down in transit depth at high metallicities.

The middle panel of Figure \ref{CompareZTP} shows the variation in the transit spectrum of our simple forward model set with temperature. There is an overall change in the shape of the spectrum and the magnitude of the spectral features caused by the temperature's effect on the scale height of the atmosphere. However, the more notable effect is that certain chemical features change dramatically with temperature. Most notably, the alkali metal lines and the CO bands grow much stronger with increasing temperature as the concentrations of these species in chemical equilibrium increase.

Finally, the bottom panel of Figure \ref{CompareZTP} shows the variation in the transit spectrum with the altitude of the cloud deck. With very deep clouds around 1000 mbar, the spectrum is nearly unaffected. As the clouds rise, they suppress the weaker features, including the Rayleigh tail and eventually produce a flat spectrum in many wavelength ranges, while the stronger spectral features are still present at a dramatically reduced amplitude.

\subsection{The Noise Model}

Our noise model for {\it JWST} assumes photon-limited noise, which is likely to be nearly correct for the relatively bright objects of interest (as opposed to read noise, which becomes important for objects much fainter than transiting exoplanet hosts). To compute the shot noise, an estimate of the stellar spectrum based on a blackbody spectrum is converted to an incoming photon count based on the star's distance and radius. This incoming photon count is multiplied by a pixel-resolution throughput function for the observing mode in question, accounting for the beam size, angular resolution, any slit losses for that mode, and the detector's quantum efficiency. The resulting spectrum is convolved with a kernel in the form of a Gaussian with a width equal to the spectroscopic resolution element to compute the number of electrons per second collected by the detector pixel by pixel.

The background model includes four thermal emission components from the telescope plus the zodiacal light, also modeled as thermal components. The zodiacal light is modeled as a blackbody with a temperature of 265 K and a directionally-varying optical depth on the order of $10^{-8}$. The thermal components are also treated as blackbodies with specific dilution factors based on how much of their emission reaches the instrument package. The four thermal components have temperatures of 38.3 K, 44.5 K, 92.0 K, and 155.0 K; and dilution factors of 0.48, 0.1, $3\times 10^{-5}$, and $9.9\times 10^{-7}$, respectively \citep{2004SPIE.5487..785S}. The backgrounds are normalized based on the slit width, if applicable. The zodiacal light is attenuated by the telescope, so it is added to the incoming stellar flux while the thermal background from the telescope is added to the count of electrons collected.

Once the number of electrons per second collected by the detector is computed, this spectrum is converted to an output signal by multiplying by the integration time and the detector efficiency. The output noise level is computed based on Poisson noise, with the noise being the square root of the electron count, and the signal being the electron count corrected to remove the zodiacal sky background. This is the output signal for the stellar flux, but for transiting planets the actual signal of interest is the difference between the in-transit and out-of-transit flux. This signal is equal to the stellar flux signal multiplied by the ratio of the planetary radius to the stellar radius squared.

\subsection{Statistical Package}

For our statistical calculations, we use the emcee Python package, an open-source MCMC code \citep{2013PASP..125..306F}. Using our theoretical forward spectral models and our noise models for the instruments, we can compute NIR and MIR spectra up to $R\gtrsim2000$, and we can also bin them to lower resolution. We can generate spectra for an arbitrary program of observing modes and observing times, and fit these synthetic spectra into a grid of model spectra of arbitrary dimension, depending upon our choice of parameterization.

Our code also allows the option of using arbitrary priors for the MCMC fit. The best possible prior is the posterior distribution from a previous observation, but this is challenging because so little is known from observational data. We address the problem of computing the effects of different priors separately in Section \ref{error}.

\subsection{Planet Selection}
\label{selection}

A number of hot jupiters have already been studied with existing observatories, providing some guidance to which objects are of greatest interest to future observations with {\it JWST} \citep{2013ApJ...779..128M,2015arXiv150407655B,2016Natur.529...59S}. These observations will be useful both to develop and refine the techniques of atmosphere characterization with higher-resolution spectra, and to address a number of unanswered questions about hot jupiters, such as the composition and structure of clouds \citep{2016ApJ...820...78L}, the temperature profiles, the possible presence of thermal inversions \citep{2007ApJ...668L.171B,2010ApJ...722..871S,2015ApJ...813...13W}, and various aspects of composition including metallicity, molecular abundances, and the atmospheric C/O ratios \citep{2013ApJ...779....3L}.

In the spirit of an observing program that must make good use of telescope time, and given the need to characterize a wide range of hot jupiters, one should model objects over a wide range of irradiation levels (a proxy in part for mean atmospheric temperature). These objects must also orbit bright stars so that they can be characterized with less total observing time. We select seven hot jupiters to model that orbit stars with magnitudes $J<11$ and equilibrium temperatures ranging from 900 K to 2500 K, all having prior transit observations, including from {\it HST}. These are HAT-P-12b, HD 189733b, HD 209458b, WASP-43b, WASP-17b, WASP-19b, and WASP-12b. With magnitudes of $J<11$ the NIRSpec Prism mode is not available, so we focus on our first five numbered observing programs, which do not use it. \footnote{For HD 189733b and HD 209458b, all of NIRSpec is unavailable, so we consider only Program 1, which does not use it at all.}

Another important parameter for the atmospheric properties is the cloudiness of the atmosphere, which does not necessarily correlate with equilibrium temperature. Therefore, to explore this dimension, we select three more hot jupiters to model from the \citet{2016Natur.529...59S} analysis that span the range of cloudiness at relatively similar equilibrium temperatures of 1100 K to 1300 K. According to their analysis, WASP-39b is the clearest of the three while WASP-6b is the cloudiest, with HAT-P-1b in the middle.

Finally, we examine the capabilities of the NIRSpec Prism mode and, concurrently, the MIRI LRS modes, which are also suitable for faint targets. Few observations have been done of planets that orbit stars faint enough to use the NIRSpec Prism, so we merely choose one representative faint target to study these modes. For this purpose, we choose Kepler-7b, which orbits a star with magnitude $J=11.83$. Important properties including those used in our forward models of our eleven selected planets are listed in Table \ref{TargetList}.

% Table 3
\begin{table*}[htbp]
\caption{Example JWST Target List}
\begin{center}
\begin{tabular}{l|r|r|r|r|r|r|r|r|l}
\hline
Planet & $T_{eq}$ & $R_{pl}$ & $M_{pl}$ & Period & Transit Duration & J-mag  & $R_*$ & $T_{eff,*}$ & Reference \rule{0pt}{2.6ex} \\ [+2pt]
           & (K)  & ($R_\Earth$) & ($M_\Earth$) & (d)   & (h)   &        & ($R_\odot$) & (K)  &      \rule[-1.2ex]{0pt}{0pt} \\ [+2pt]
\hline
WASP-12b   & 2548 & 1.736        & 1.404        & 1.091 & 3.008 & 10.477 & 1.599       & 6300 & (1)       \rule{0pt}{2.6ex}  \\ [+2pt]
WASP-19b   & 2091 & 1.395        & 1.114        & 0.789 & 1.668 & 10.911 & 1.004       & 5500 & (2)                          \\ [+2pt]
WASP-17b   & 1661 & 1.991        & 0.486        & 3.735 & 3.743 & 10.509 & 1.380       & 6650 & (3)                          \\ [+2pt]
WASP-43b   & 1442 & 1.036        & 2.034        & 0.813 & 1.210 &  9.995 & 0.667       & 4520 & (4)                          \\ [+2pt]
HD 209458b & 1440 & 1.380        & 0.714        & 3.525 & 3.024 &  6.591 & 1.146       & 6075 & (5)                          \\ [+2pt]
HD 189733b & 1203 & 1.138        & 1.138        & 2.219 & 1.869 &  6.070 & 0.781       & 4980 & (6)                          \\ [+2pt]
HAT-P-12b  &  958 & 0.959        & 0.211        & 3.213 & 2.085 & 10.794 & 0.701       & 4650 & (7)  \rule[-1.2ex]{0pt}{0pt} \\ [+2pt]
\hline
WASP-39b   & 1116 & 1.270        & 0.280        & 4.055 & 2.800 & 10.663 & 0.895       & 5400 & (8)       \rule{0pt}{2.6ex}  \\ [+2pt]
HAT-P-1b   & 1294 & 1.217        & 0.524        & 4.465 & 2.419 &  9.156 & 1.115       & 5975 & (9)                          \\ [+2pt]
WASP-6b    & 1195 & 1.224        & 0.503        & 3.361 & 2.738 & 10.769 & 0.870       & 5450 & (10) \rule[-1.2ex]{0pt}{0pt} \\ [+2pt]
\hline
Kepler-7b  & 1628 & 1.614        & 0.433        & 4.886 & 5.616 & 11.833 & 2.020       & 5933 & (11) \rule{0pt}{2.6ex} \rule[-1.2ex]{0pt}{0pt} \\ [+2pt]
\hline
\end{tabular}
\end{center}
\tablerefs{
(1) \citet{2009ApJ...693.1920H};
(2) \citet{2010ApJ...708..224H};
(3) \citet{2010ApJ...709..159A};
(4) \citet{2011A&A...535L...7H};
(5) \citet{2000ApJ...532L..51C};
(6) \citet{2005A&A...444L..15B};
(7) \citet{2009ApJ...706..785H};
(8) \citet{2011A&A...531A..40F};
(9) \citet{2007ApJ...656..552B};
(10) \citet{2009A&A...501..785G};
(11) \citet{2010ApJ...713L.140L}
}
\label{TargetList}
\end{table*}

Our individual spectral models for these eleven modeled planets are loosely based on the available observations for each object, with some adjustments to place them at the midpoints between the grid points in order to present a worst-case scenario for our retrieval algorithm. For all eleven objects, we set the metallicity to +0.75 dex, based on the metallicity of Jupiter's atmosphere of +0.5 dex \citep{2004Icar..171..153W} increased to midway between grid points. We set the temperature to the equilibrium temperature, rounded up to the nearest midpoint between grid points. For the seven objects we selected to span the temperature range, as well as Kepler-7b, we set the cloud pressure level to roughly 133 mbar. This is a higher altitude than the cloud deck at several hundred millibars in Jupiter's (much colder) atmosphere \citep{2004jpsm.book...79W}, but it is lower than the cloud decks at $\lesssim$10 millibars observed in some hot exoplanet atmospheres \citep{2015MNRAS.446.2428S,2016ApJ...817..141S}. We set the pressure nearer to the lower value so that we can model cloudier atmospheres with plausible higher cloud decks.

\citet{2016Natur.529...59S} use the difference in altitude corresponding to their UB-LM spectral index in transit depth, divided by the scale height of the atmosphere, as a proxy for cloudiness. For WASP-39b, they find a difference in altitude of $0.10\pm 0.41$ scale heights. For HAT-P-1b, they find a difference of $2.01\pm 0.81$ scale heights, and for WASP-6b, they find a difference of $8.49\pm 1.33$ scale heights. These spectral indices correspond to a hazy atmosphere rather than a cloudy one, so they do not readily lend themselves to a single cloud pressure level, as in our models. Therefore, we somewhat arbitrarily set their cloud pressures at 133 mbar, 13.3 mbar and 1.33 mbar, respectively.

However, our analysis finds that, at least for our three-parameter model, the scale height of the atmosphere and the brightness of the target are the two most important planet parameters. A larger scale height results in larger transit features and a greater amount of mutual information obtained, while for the target brightness, there are three distinct ranges determined by which instruments and modes can be used to observe them. Bright objects with $J<8$ include HD 189733b and HD 209458b and are observable with NIRISS, NIRCam, and MIRI LRS. Intermediate brightness objects with $8<J<11$ include all of our selected HATNet and WASP objects and are observable with all of the modes except the NIRSpec Prism. Faint objects with $J>11$ include Kepler-7b and are observable only with the NIRSpec Prism and MIRI LRS.

% % % % % % % % % % % % % % % % % % % % % % % % % % % % % % % % % % % % % % % % % % % % % % % % % % % % % % % % % % % % % % % % % % % % % % %

\section{Results}
\label{results}

%We base most of our results on an in-depth analysis of observing programs for HAT-P-1b. We choose this object because it is the brightest object in our sample that is observable in all four instruments. We explore other objects in Sections \ref{modes}-\ref{faint}.

%\vspace{5cm}
\subsection{Jacobians}

\begin{figure*}[htp]
\includegraphics[width=0.99\textwidth]{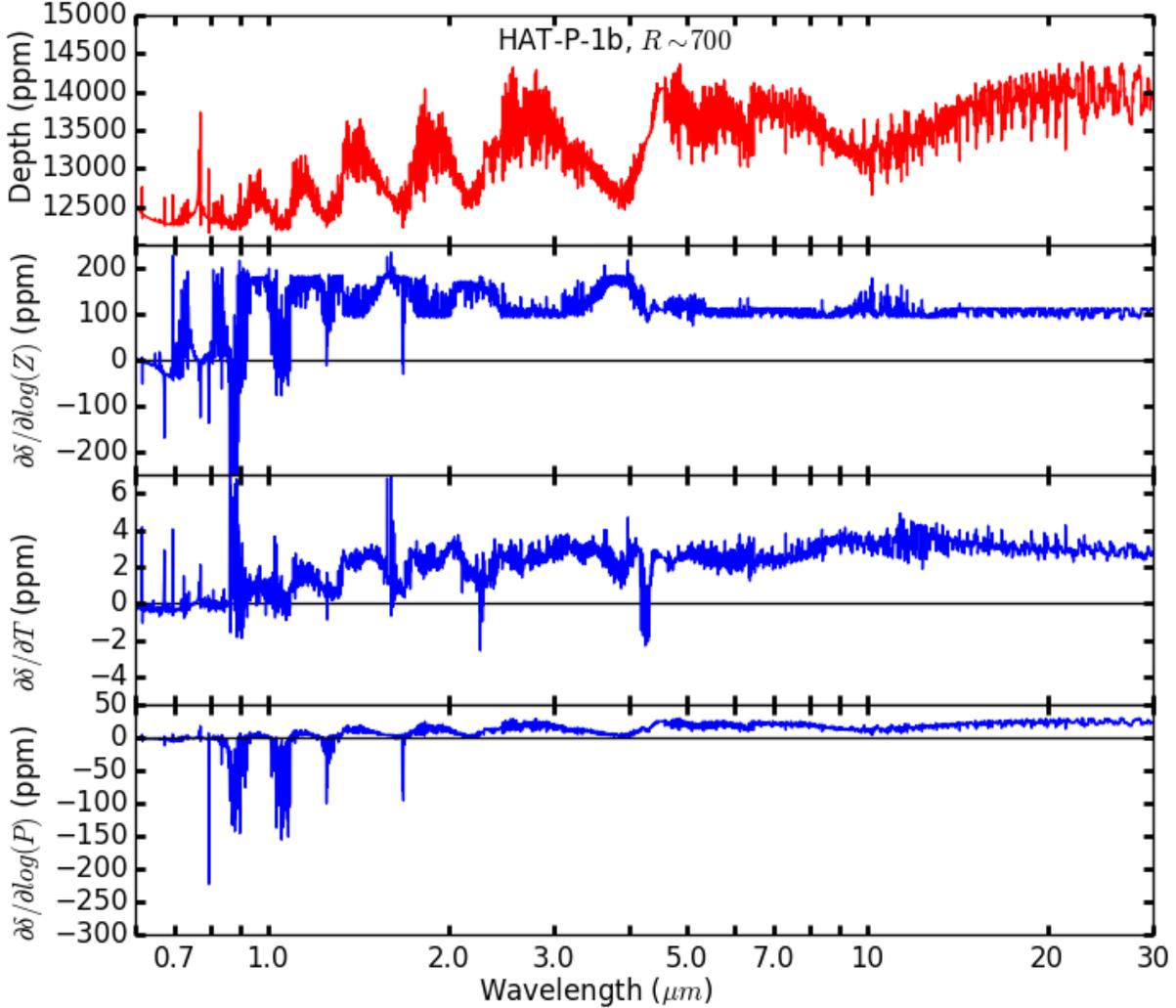}
\caption{Transit depth of HD 189733b plotted with the components of the Jacobian with respect to the model parameters, at a medium resolution of R$\sim$700. Large excursions from zero indicate a strong dependence of the transit depth on that variable at that wavelength. Thus, observing at those wavelengths will provide strong diagnostics for those parameters.}
\label{jacobiansM}
\end{figure*}

For demonstration purposes, Figure \ref{jacobiansM} shows the model transit spectrum of HAT-P-1b, a relatively bright target observable with all four instruments, along with the components of the Jacobian of the transit spectra with respect to our three model parameters, log(Z), T, and log(P), where Z, T, and P are the metallicity, atmospheric temperature, and pressure at the cloud tops, respectively. The Jacobians are particularly useful in that they give guidance about the wavelength ranges that are most sensitive to each parameter. The plot indicates that the spectrum's sensitivity to metallicity is relatively uniform across the entire frequency range, but especially so in the mid-IR range probed by MIRI LRS.

The most sensitive bands to temperature are several alkali metal lines at short wavelengths such as the potassium resonance line at 0.77 microns, and the wings of the water absorption features at 0.97 and 1.95 microns. The excursions from positive to negative Jacobians in the troughs of the transit spectrum could also provide useful diagnostics. The spectrum is most sensitive to the cloud pressure level in the near-IR around 1 $\mu m$, specifically, in the gaps between the water bands$-$bands with low molecular opacities that are ``filled in'' by obscuring clouds. Note, however, that with our simple, three-parameter model, we do not need to make use of the Jacobians to narrow our search for an optimal observing program because we can explore all of the possibilities throughout the parameter space of the model set.

\subsection{Code Verification}
\label{verify}

\begin{figure*}[htp]
\includegraphics[width=0.99\textwidth]{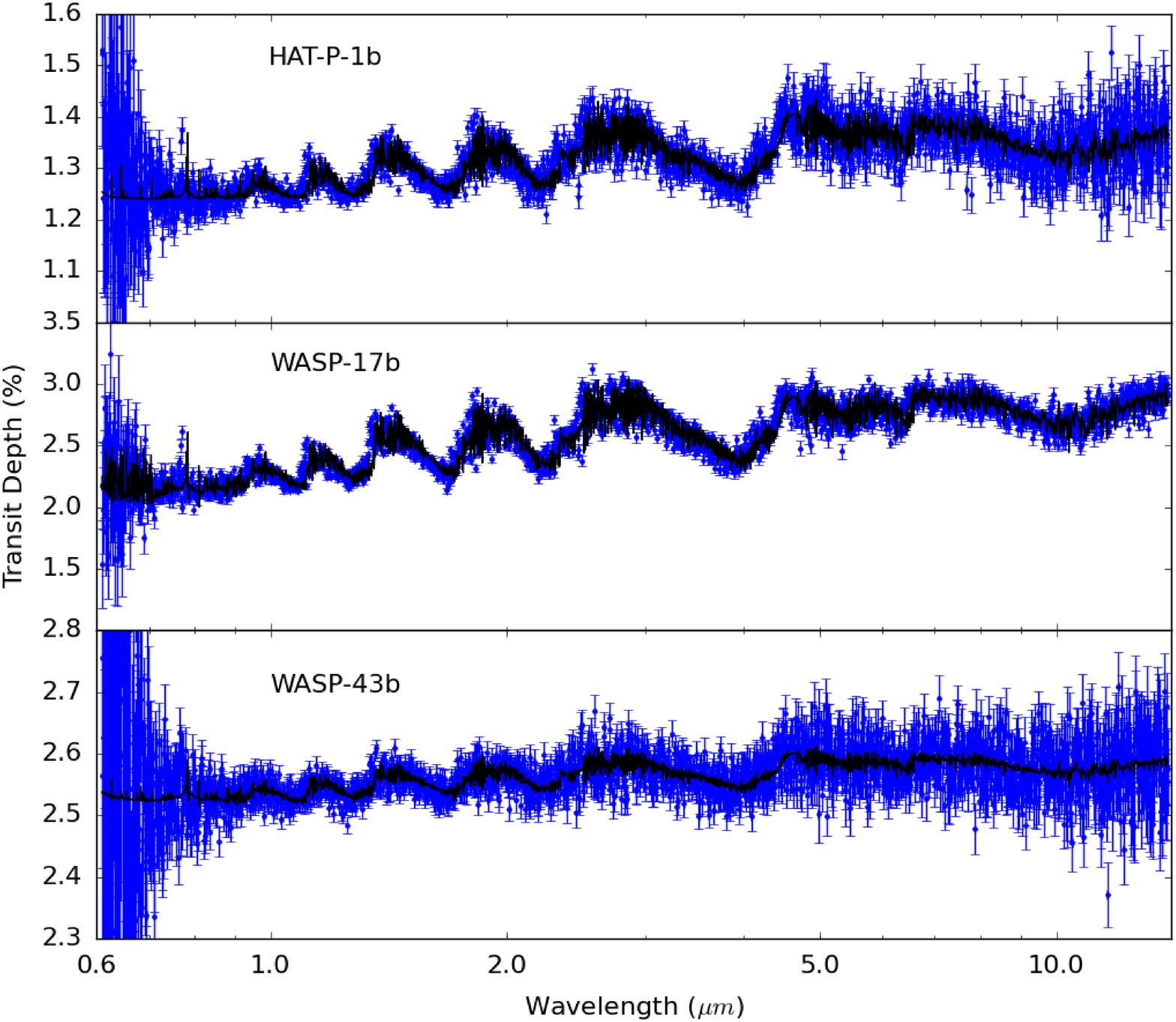}
\caption{Best-fit spectra to synthetic observations of HAT-P-1b, WASP-17b, and WASP-43b generated by our retrieval code. The fits are made to 10 hours of total in-transit observations with our Program 1. WASP-17b has the strongest transit signature in our sample, while WASP-43b has the weakest.}
\label{PlanetFits}
\end{figure*}

\begin{figure*}[htbp]
	\subfigure{\includegraphics[width=0.40\textwidth]{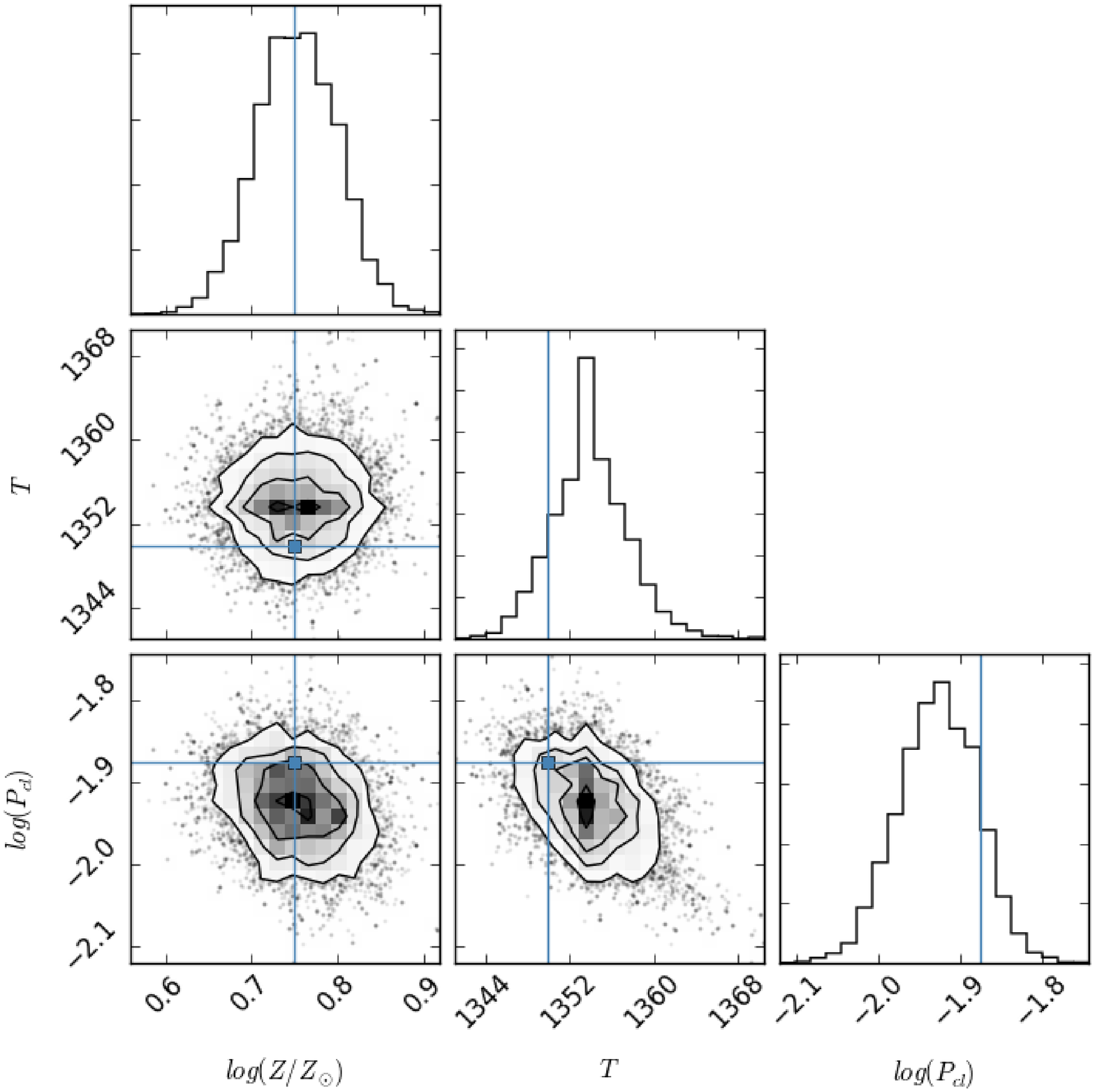}}
	\subfigure{\includegraphics[width=0.40\textwidth]{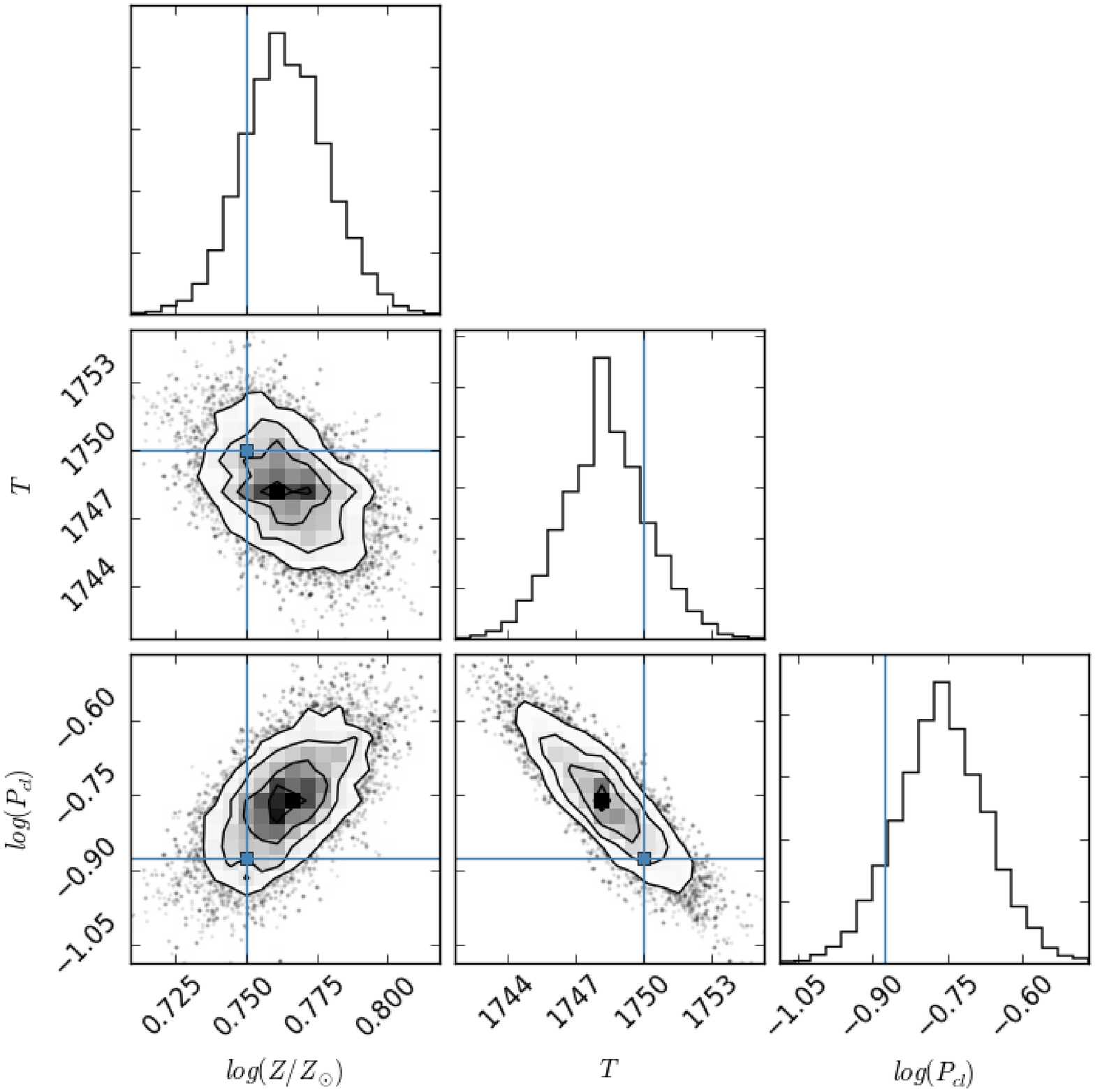}}
	\subfigure{\includegraphics[width=0.40\textwidth]{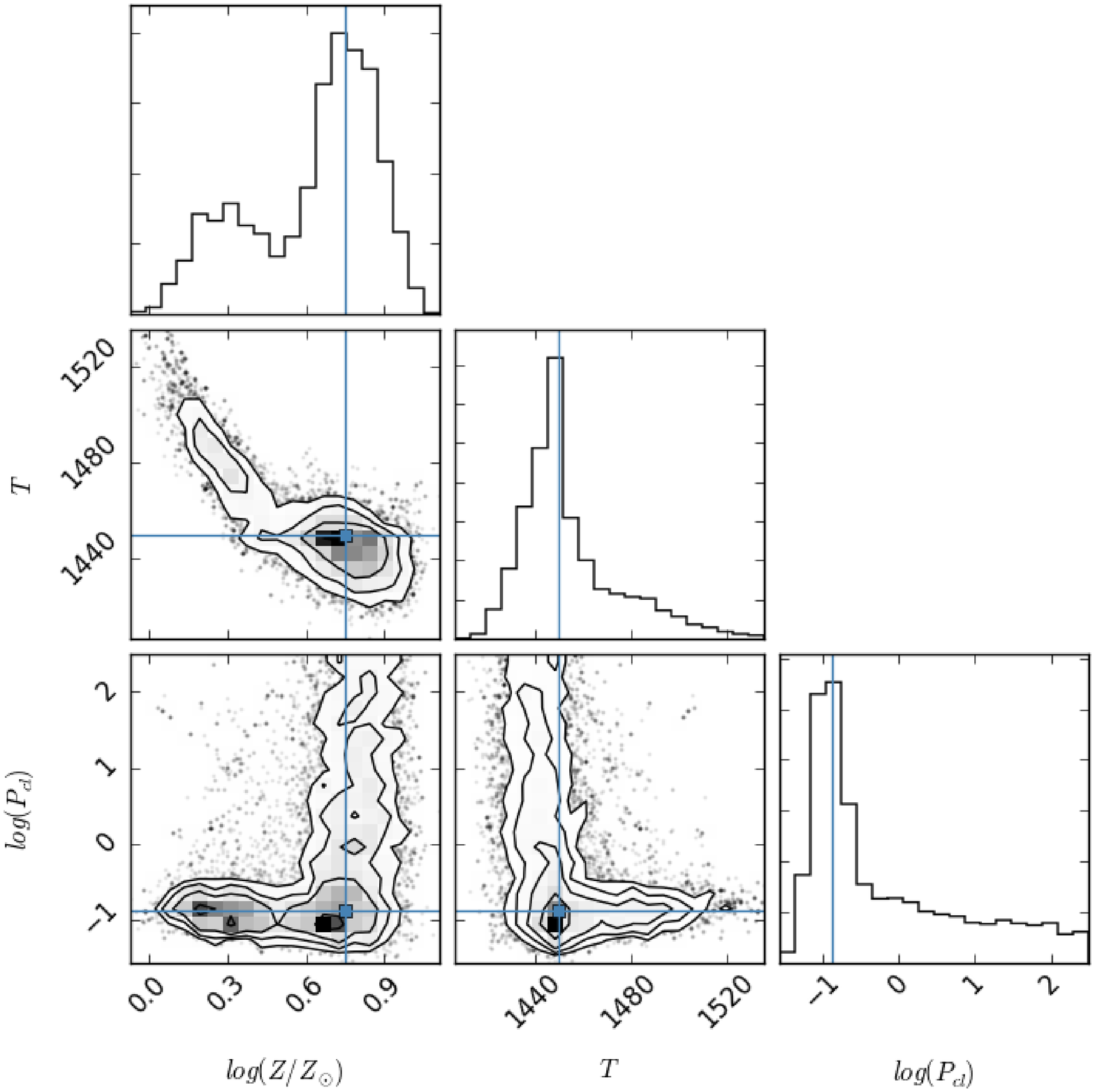}}
	\caption{Posterior distributions from retrievals of our three model parameters for synthetic observations of HAT-P-1b (top left), WASP-17b (top right), and WASP-43b (bottom left). Each observation uses 10 hours of total in-transit observing time for Program 1.}
	\label{SpecDistr}
\end{figure*}

For a given forward model set and synthetic spectrum, our retrieval code outputs a best fit of the model parameters and a posterior distribution. To verify the accuracy of our code, we show the spectral fits and retrievals for three planets: HAT-P-1b, WASP-17b, and WASP-43b in Figures \ref{PlanetFits} and \ref{SpecDistr}, respectively. We perform these fits for a total of 10 hours of in-transit observation for Program 1 for each planet. These three planets span the range of strength of transit signatures due the different scale heights of their atmospheres (see Section \ref{individual}).

WASP-17b has the strongest transit signature of any planet in our sample, resulting in proportionately smaller error bars and a better spectral fit. WASP-43b, on the other hand, has the weakest transit signature, proportionately larger error bars and a poorer fit. HAT-P-1b is roughly in the middle However, all three spectral fits follow the mean of the synthetic observations quite well. We also note that the error bars become much larger blueward of 0.8 microns, since the telescope is not optimized for such short-wavelength observations.

The posterior distributions are plotted in triangle plots with 1- and 2-dimensional histograms of the probability density function marginalized over each of our model parameters. The retrievals recover the input parameters quite well, with uncertainties on the order of 10 K in temperature and on the order of 0.1 dex in metallicity and cloud top pressure for HAT-P-1b. For comparison, for WASP-17b, we recover the metallicity and temperature at about twice as great a precision, but the cloud top pressure is recovered more poorly, with an uncertainty of $\sim$0.2 dex.

On the other hand, the retrieval for WASP-43b is considerably less precise, likely because the transit signature of the planet smaller. The uncertainties in all three parameters are significantly greater, and the posterior distribution is bimodal with two possible solutions: one near the correct solution of $log(Z/Z_\odot)=0.75$, $T=1450$ K, and $P_cl=133$ mb, and one at roughly $log(Z/Z_\odot)=0.25$, $T=1480$ K, and $P_cl=133$ mb. This is likely due to a degeneracy between the spectra of hotter, more metal-poor atmospheres and cooler, more metal-rich atmospheres, since temperature and metallicity have opposite effects on the strength of the transit signature. Our code is also much poorer at recovering $P_{cl}$ for this object, leaving it effectively unconstrained at 2-$\sigma$ confidence, even though the best fit solution is still correct. This is mainly because increasing $P_{cl}$ pushes the clouds deeper in the atmosphere, where they have very little effect on the transit spectrum.

Thus, we see that the retrieval code functions as expected over a range of planet parameters, although it does not cover all parameters equally, with $P_{cl}$ being the weakest of the three in our model. This is likely because the various parameters are best recovered at different wavelengths and in different observing modes.

\subsection{Estimates of Mutual Information and Priors}
\label{error}

The mutual information between the prior and posterior distributions of the model parameters for a given observing program is approximately the difference in their entropies. However, the entropy of the posterior is actually the {\it average} entropy of the posterior for a particular observation. To do a full Monte Carlo integration of the mutual information is computationally very expensive, so we use the entropy of the posterior for a single MCMC simulation as an estimate. We can estimate the actual mutual information to greater precision and the typical error of the single MCMC run by doing a very low-resolution sampling of the Gaussian distribution of possible observations and taking the average. We have done this by computing 1000 posterior distributions for the model parameters based on a 1.0-hour integration of HAT-P-1b with the NIRCam F322W2 filter, each with a different random seed according to the Gaussian likelihood function. We then compute the entropy for each distribution and take an average.

It is possible to estimate priors for our retrieval algorithm based on existing transit spectroscopy data, such as those from {\it HST}, provided the conclusions drawn from them can be justified. If no such spectroscopic data are available, however, the options are far more limited. The properties of a planet's orbit and host star allow us to estimate an equilibrium temperature, but even this is subject to uncertainties in the planetary albedo. Planetary mass appears to have some relation to atmospheric metallicity, but with significant scatter \citep{2016ApJ...831...64T}. Most other parameters are effectively unconstrained {\it a priori}.

We explore the effect of the choice of priors on the results by comparing two simple priors: a flat prior over the full parameter space of our model set and a broad Gaussian prior centered around the estimated equilibrium temperatures and metallicities of the planets with $\sigma_{\log Z} = 0.5$, $\sigma_T = 200$ K, and $\sigma_{\log P} = 1.0$. Note that each of these priors has a different amount of information inherent, or assumed, in it, which must be justified when comparing the results. To show this difference, we can compute the entropies of our two priors directly.

In the units we use, the multivariate Gaussian prior has $H = \log_2(\sigma_{\log Z}\sigma_T\sigma_{\log P}(2\pi e)^{3/2}) = 12.79$ bits. We can similarly calculate the entropy per degree of freedom for this prior: $H/N = 4.26$ bits. Meanwhile, for the uniform prior, $H/N = \log_2((\log Z_{max}-\log Z_{min})(T_{max}-T_{min})(\log P_{max}-\log P_{min}))/3 = 4.66$ bits, if $\Delta T$ (which can vary between planets) is 1000 K. All of these values are unit-dependent, but the difference, $[H(flat)-H(Gauss)]/N = 0.40$ bits, is dimesionless and is thus our figure of interest. In other words, the Gaussian prior assumes 0.40 bits more information per degree of freedom than the uniform prior.

\begin{figure}[htp]
\includegraphics[width=\columnwidth]{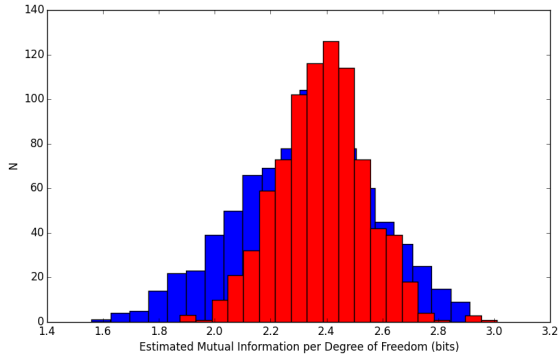}
\caption{Histograms of estimates of mutual information per degree of freedom for observations of HAT-P-1b with the NIRCam F322W2 mode, relative to a uniform prior. Blue is the covariance method, and red is the posterior entropy method.}
\label{histplot}
\end{figure}

We estimate the mutual information provided by a given observation using our covariance matrix method and our posterior entropy method described in Section \ref{info}. To determine how accurate these methods are, we compute 1000 MCMC simulations of the same observation in this case, a 1-hour observation of HAT-P-1b with the F322W2 mode of NIRCam, sampled from a Gaussian distribution, each of which gives different estimates of the mutual information. We plot a histogram of these estimates for both methods in Figure \ref{histplot}, the covariance matrix method in blue and the posterior entropy method in red. If the posterior distributions were Gaussian, the histogram would be a delta function, and we would know the mutual information exactly. As they are non-Gaussian, the set of estimates itself has a distribution (which is narrower for the posterior entropy method), which tells us the root-mean-square error of our approximation methods.

For a flat prior, our distribution of estimates of $I(\Theta,X)/N$ as computed with the posterior entropy method, has a mean of 2.38 bits and a standard deviation of 0.16 bits. For a Gaussian prior, with the same method, we find a mean of 1.98 bits and a standard deviation of 0.15 bits. With both priors, we can obtain accurate estimates of the mutual information per degree of freedom within $\sim$0.3 bits to $2\sigma$ confidence, and the difference between the mutual information per degree of freedom in the two cases is almost exactly equal to the 0.4-bit difference in the entropy of the priors. In other words, the flat prior is statistically more informative because it assumes less, and the error bars are nearly independent of the choice of prior. (This can be seen because the error bars of the posterior distribution depend on the posterior entropy, which shows almost zero difference between the two priors.)

When we use the same procedure with the covariance matrix method of estimating the mutual information per degree of freedom, we get a different distribution of estimates. When a flat prior is used, we find a distribution with a mean of 2.32 bits and a standard deviation of 0.25 bits, but when a Gaussian prior is used, we find a mean of 1.82 bits and a standard deviation of 0.23 bits. Based on these numbers, the flat prior is still more informative, but by an even greater a margin of 0.50 bits. In other words, the difference in the mutual information per degree of freedom is {\it greater} than the difference in the entropies of the prior. On its face, this would suggest that the ``broader'' flat prior might constrain the posterior distribution better, although the gain is small (only 0.1 bits), so it may not be significant. Either way, though, the flat prior is statistically more informative than the Gaussian in this case. This conveys no advantage if this prior cannot be justified, and the best choice of prior is likely to be the posterior for a previous study. However, between our two test priors, we use the flat prior for the remainder of our analysis.

Our posterior entropy method of estimation appears to compute the mutual information more accurately based on the narrower width of the distribution of estimates and the nearly-identical peak location, so we feel confident enough to use this method for the remainder of our analysis. We still estimate the mutual information using both methods, but in several plots, we plot only the results from the posterior entropy method.

\subsection{MIRI LRS: Slit v. Slitless}
\label{LRS}

The other choice we must make for our analysis is whether to use the Slit or Slitless mode for MIRI LRS, and we can choose it in the same way. While we do not take an average of many runs of the same observation, we find that the difference between the mutual information yielded by the Slit and Slitless modes is fairly consistent between different observing modes and even different objects. We can even look at the statistical distribution of these results to determine whether one mode is consistently more informative than the other.

We estimated the mutual information for each observation using the posterior entropy method, and the estimated mutual information for the Slitless mode was consistently greater than the Slit mode for all of our model observations. The average of these results was 2.57 bits per degree of freedom for the Slit mode and 3.53 bits for the Slitless mode. Thus, the Slitless mode yielded and average of 0.97 bits more information, which is reasonable given the higher throughput of the Slitless mode. The standard deviation of this difference was 0.35 bits. Notably, the Slitless mode provides the greatest advantage for WASP-12b and the smallest advantage for HAT-P-1b.

Estimating the mutual information by the covariance matrix method produced similar results. In only 1 out of 30 observations was the Slitless mode less informative than the Slit mode. The average mutual information yielded by the Slit mode was 2.69 bits, while for the Slitless mode, it was 3.70, for a difference of 1.01 bits, with a standard deviation of 0.49 bits.

Theoretically, the Slitless mode has greater throughput and relatively low noise levels for bright objects like exoplanet hosts, which should make it better for planet observations in general, which is supported by our results for our particular model. Therefore, we use the Slitless mode in our numbered observing programs and our other subsequent analyses.

\subsection{Integration Time Dependence}
\label{modes}

We computed posterior distributions for all of {\it JWST's} observing modes and for all of our numbered observing programs for each of the planets on our list for a range of integration times. We used these to estimate the mutual information obtained with each observing mode. In particular, we computed total in-transit observing times from 1.0 to 10.0 hours (total for all the visits in an observing program) in logarithmic steps of $10^{0.2}\approx 1.6$. This metric of in-transit integration time is somewhat incomplete because out-of-transit observations of the star will also be needed to establish a baseline spectrum. Our noise model also does not account for the advantage of transit light curve fitting of a whole transit compared with a partial transit (e.g. observing all of a 1-hour transit may be more informative than half of a 2-hour transit because we can observe the shape of the light curve across the spectrum and fit it better to the baseline). However, for most of the observations we model in this paper, each visit will cover only a fraction of a transit, so this issue will have less effect on our study of integration time by itself. However, in a planned observing program, it will be necessary to choose observing times carefully based on the operations of the telescope and actual transit durations of the targets in order to acquire full transits, when possible, as well as the needed out-of-transit observations.

In considering the effect of observing time alone, as a check, we can predict the effect of increasing observing time (adding multiple transits when necessary) by noting that $\sigma \sim 1/\sqrt{t_{obs}}$ to first order and therefore, increasing the observing time by a factor of 1.6 should result in a gain on the order of $\Delta I/N \sim -\log_2(10^{-0.1}) = 0.33$ bits of information. (The factor of N accounts for the 3 model parameters.)

Table \ref{Info} show lists the estimated mutual information per degree of freedom for the eight planets that fall within our medium brightness range, as observed with Program 1. We also include first-order estimates of the scale heights of the atmospheres, which we explore in Section \ref{individual}. The slopes of the trendlines for the eight objects, according to a least-squares fit of the values, are quite similar. We find that increasing the observing time by a factor of 1.6 increases the mutual information per degree of freedom by an average of 0.32 bits when estimated with the posterior entropy method and by 0.31 bits per step when estimated with the covariance matrix method, with little variation between objects. Both of these methods are very close to our $1/\sqrt{t}$ first-order estimate. This is to be expected given our approximation of Poisson noise in our noise model, although it is encouraging that it is not affected by our thermal and sky backgrounds.

This result is notable in that {\it JWST} does not reach diminishing returns in terms of mutual information even after a total in-transit integration time over multiple visits of 10 hours. This might be the case for fainter objects for which noise levels are greater, in which case it might be of less use to observe longer, but for exoplanet hosts this is not an issue, and the primary restrictions on observing time are the allocated project time, the transit time and thus the number of visits required, and the desired level of precision.

\begin{figure}[htp]
\includegraphics[width=\columnwidth]{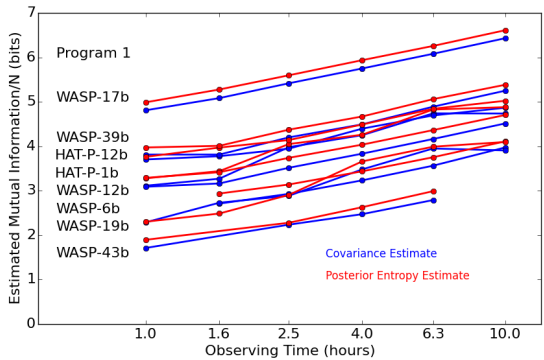}
\caption{Estimated mutual information per degree of freedom versus observing time for eight modeled planets for our Observing Program 1. Both the covariance matrix (blue) and posterior entropy (red) methods of estimation are shown.}
\label{InfoTimeFig}
\end{figure}

The mutual information per degree of freedom estimated with the posterior entropy method (which we found to be more precise) is systematically higher than that calculated by the covariance matrix method, albeit by a small amount, less than one bit in most cases. This is easier to see in Figure \ref{InfoTimeFig}, which plots these values visually. Here, the blue lines represent the covariance matrix estimates, and the red lines represent the posterior entropy estimates.

\begin{table*}[htbp]
\caption{Estimates of Mutual Information Per Degree of Freedom for Observing Program 1}
\begin{center}
\begin{tabular}{l|l|r|r|r|r|r|r|r}
\hline
Object    & Estimate   & Scale Height (km) & 1.0 h & 1.6 h & 2.5 h & 4.0 h & 6.3 h & 10.0 h \\
\hline
HAT-P-1b  & Covariance &  524 &  3.28 &  3.44 &  4.05 &  4.26 &  4.82 &  4.88  \\
HAT-P-1b  & Entropy    &  524 &  3.11 &  3.27 &  3.98 &  4.25 &  4.74 &  4.74  \\
HAT-P-12b & Covariance &  598 &  3.76 &  3.97 &  4.14 &  4.49 &  4.84 &  5.02  \\
HAT-P-12b & Entropy    &  598 &  3.70 &  3.78 &  3.95 &  4.40 &  4.69 &  4.87  \\
WASP-6b   & Covariance &  510 &       &  2.94 &  3.14 &  3.44 &  3.75 &  4.11  \\
WASP-6b   & Entropy    &  510 &       &  2.71 &  3.93 &  3.23 &  3.56 &  3.97  \\
WASP-12b  & Covariance &  783 &  3.29 &  3.42 &  3.74 &  4.04 &  4.36 &  4.70  \\
WASP-12b  & Entropy    &  783 &  3.09 &  3.16 &  3.52 &  3.83 &  4.16 &  4.51  \\
WASP-17b  & Covariance & 1940 &  4.99 &  5.28 &  5.60 &  5.94 &  6.26 &  6.61  \\
WASP-17b  & Entropy    & 1940 &  4.81 &  5.08 &  5.42 &  5.75 &  6.08 &  6.43  \\
WASP-19b  & Covariance &  523 &  2.30 &  2.49 &  2.91 &  3.66 &  3.99 &  4.10  \\
WASP-19b  & Entropy    &  523 &  2.29 &  2.73 &  2.89 &  3.48 &  3.95 &  3.91  \\
WASP-39b  & Covariance &  921 &  3.97 &  4.01 &  4.37 &  4.67 &  5.06 &  5.38  \\
WASP-39b  & Entropy    &  921 &  3.81 &  3.81 &  4.20 &  4.49 &  4.89 &  5.25  \\
WASP-43b  & Covariance &  109 &  1.90 &       &  2.28 &  2.63 &  2.99 &        \\
WASP-43b  & Entropy    &  109 &  1.71 &       &  2.24 &  2.47 &  2.79 &        \\
%Average   & Covariance &  738 &  3.36 &  3.65 &  3.78 &  4.14 &  4.51 &  4.97  \\
%Average   & Entropy    &  738 &  3.22 &  3.51 &  3.64 &  3.99 &  4.36 &  4.81  \\
\hline
\end{tabular}
\end{center}
\label{Info}
\end{table*}

\subsection{Individual Planets}
\label{individual}

Many factors could potentially impact the mutual information obtained for individual planets including equilibrium temperature and cloudiness, the two parameters we used to select our target list. The target brightness could also be important, since brighter targets would give better statistics. However, to first order, the largest effect should be due to the scale height of the atmosphere. With everything else being equal, doubling the scale height would also double the magnitude of the transit features, which will double the signal-to-noise ratio for the same noise level. Again, to first order, this would reduce the error bars of the retrieved model parameters by a factor of 2, adding 3 bits of information across the three model parameters, or 1 bit per degree of freedom.

The scale height is $H_{sc} = kT/\mu g$, to first order, where $\mu$ is the mean molecular weight, and $g$ is the surface gravity. This means that $H_{sc}\propto T_{eq}R^2/M$. We include this first-order estimate of the scale height of each planet in Table \ref{Info} to show the correlation between scale height and mutual information per degree of freedom. The planet with the largest scale height is WASP-17b, while the planet with the smallest scale height is WASP-43b. Based on the values in Table \ref{TargetList}, the ratio of their scale heights is approximately 17.7, and based on this first-order estimate, identical observations should yield $\log_2(17.7) = 4.15$ bits more mutual information per degree of freedom for WASP-17b than for WASP-43b. However, this is a maximum estimate assuming no correlation between the model parameters, which is not the case in our simulations, so the actual value may be smaller. Indeed, the actual difference between the mutual information per degree of freedom obtained for these two objects in our simulations is an average of 3.19 bits.

We also see that WASP-6b and WASP-19b yield systematically less mutual information that the other objects besides WASP-43b, and they also have smaller estimated scale heights, although not by much. There is little correlation between the mutual information and scale height for the remaining four objects, but without knowing any significant information about the atmospheres {\it a priori}, this is not unreasonable because differences in composition, thermal structure, or cloud cover could account for the difference.

One metric that would be useful is the amount of observing time needed to quantify a measurement to a specified level of precision. This is different for each object, and to compute it involves the transit depth and scale height of the particular planet and also relies on the model used. However, in most of our calculations, the error bars still obey the $\sigma \sim 1/\sqrt{t}$ rule of thumb for times likely to be used in an observing campaign, and we can estimate the amount of observing time needed for these objects based on our simple forward model from our results in this study. This is shown in Table \ref{InfoTime}. For this table, we compute the amount of observing time needed to obtain 12 bits of mutual information, or 4 bits per degree of freedom.

\begin{table}[htbp]
\caption{Integration Time to Obtain 4 Bits per Degree of Freedom with Observing Program 1}
\begin{center}
\begin{tabular}{l|r}
\hline
Object    & Observing Time (h)\\
\hline
HAT-P-1b  &  4.59 \\
HAT-P-12b &  2.93 \\
WASP-6b   & 10.14 \\
WASP-12b  &  6.65 \\
WASP-17b  &  0.46 \\
WASP-19b  & 12.85 \\
WASP-39b  &  2.36 \\
WASP-43b  & 72.72 \\
\hline
\end{tabular}
\end{center}
\label{InfoTime}
\end{table}

The scale height of a planet's atmosphere makes an enormous difference to the required observing time because the mutual information is strongly correlated with the scale height, and the observing time needed increases exponentially with the mutual information demanded. For most of the objects, a few hours are needed to reach 12 bits, but for WASP-17b, only half an hour is needed to get the same results, while for WASP-43b, the requirement is a probably unrealistic 73 hours. There is thus a great advantage in focusing an observing program on the less dense, ``puffy'' hot jupiters, which have large scale heights, while some objects, like WASP-43b would be much more difficult to take useful observations.

\subsection{Observing Modes and Programs}

\begin{figure*}[htbp]
\subfigure{\includegraphics[width=0.40\textwidth]{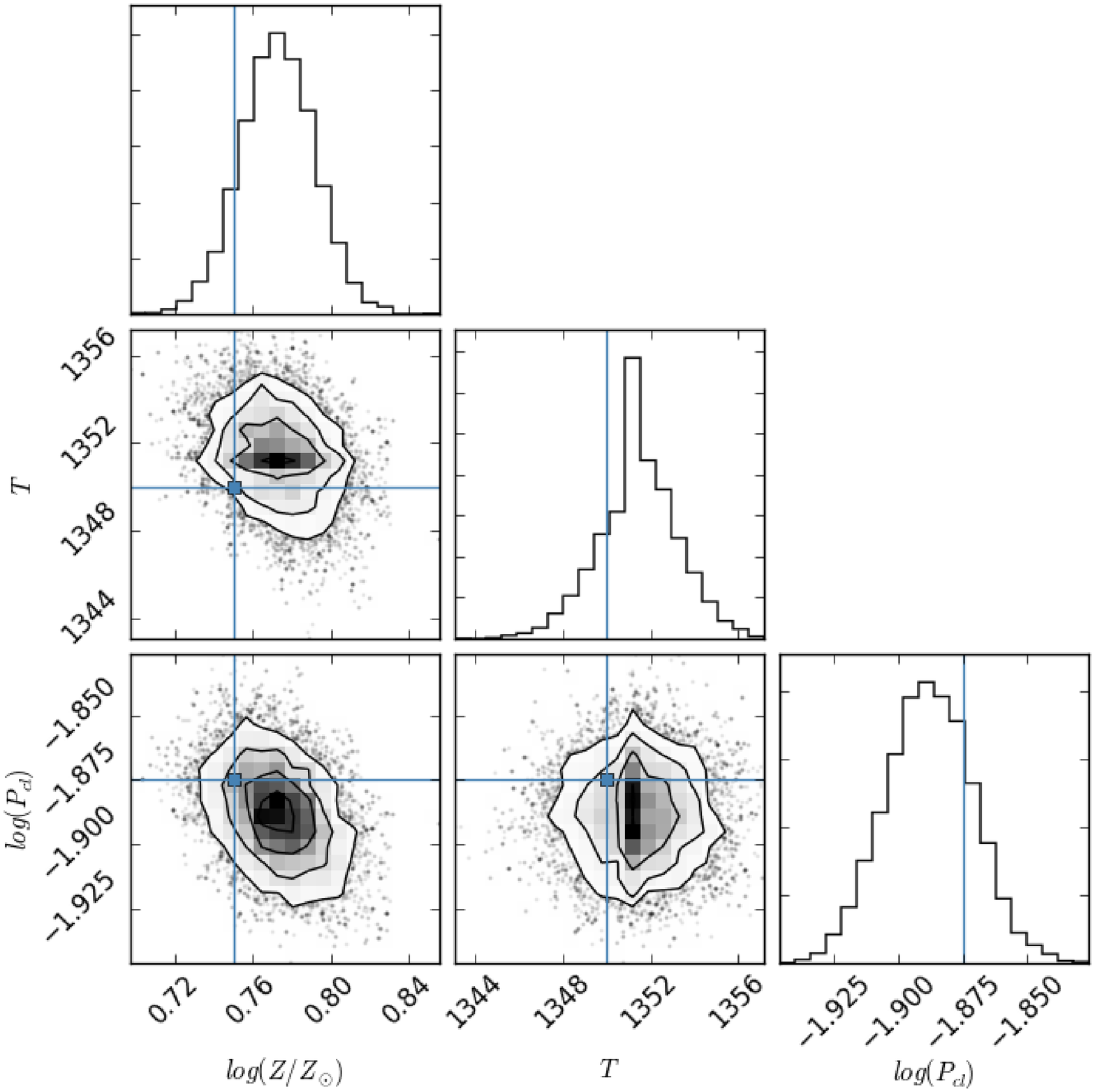}}
\subfigure{\includegraphics[width=0.40\textwidth]{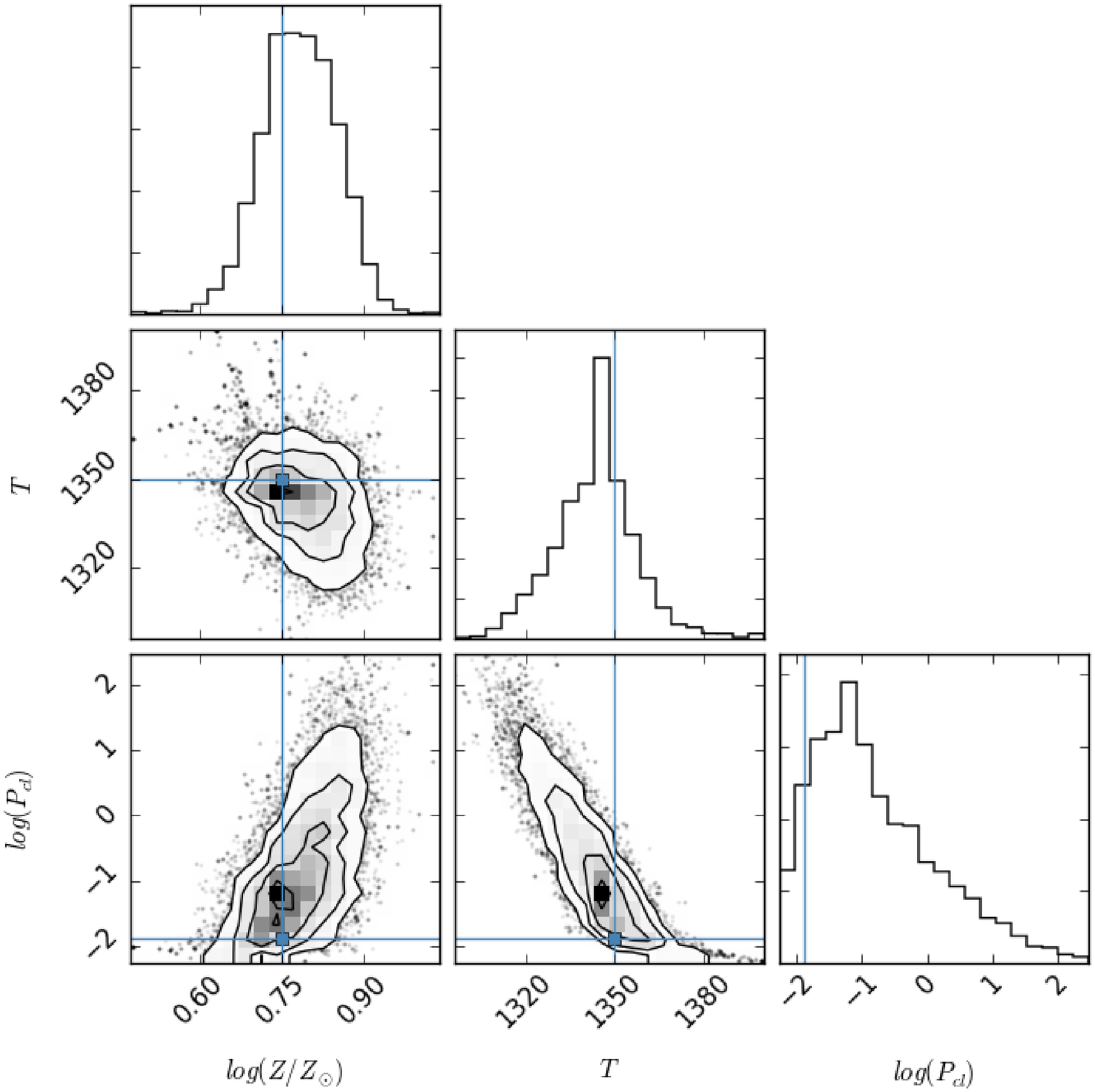}}
\subfigure{\includegraphics[width=0.40\textwidth]{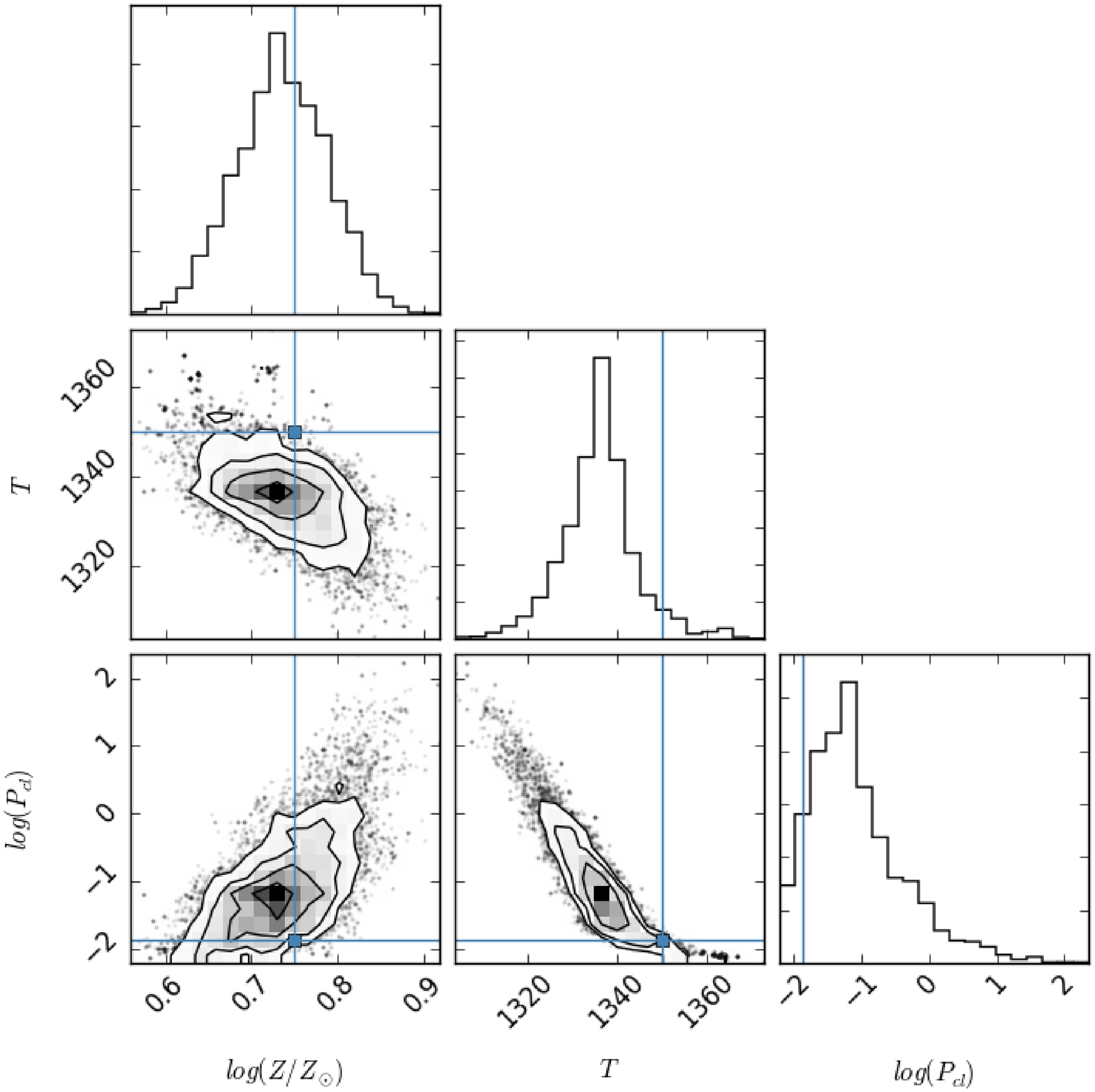}}
\subfigure{\includegraphics[width=0.40\textwidth]{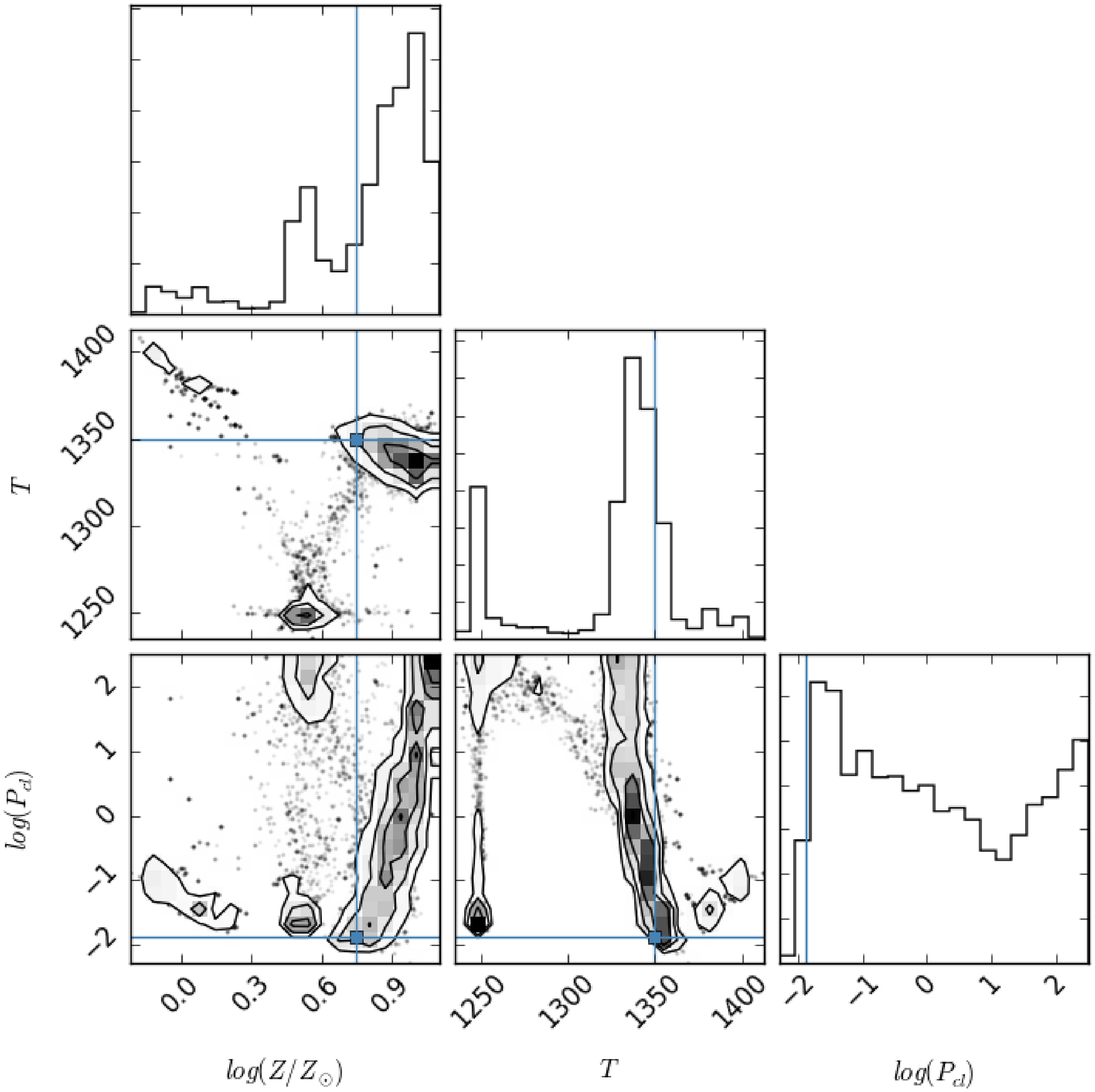}}
\subfigure{\includegraphics[width=0.40\textwidth]{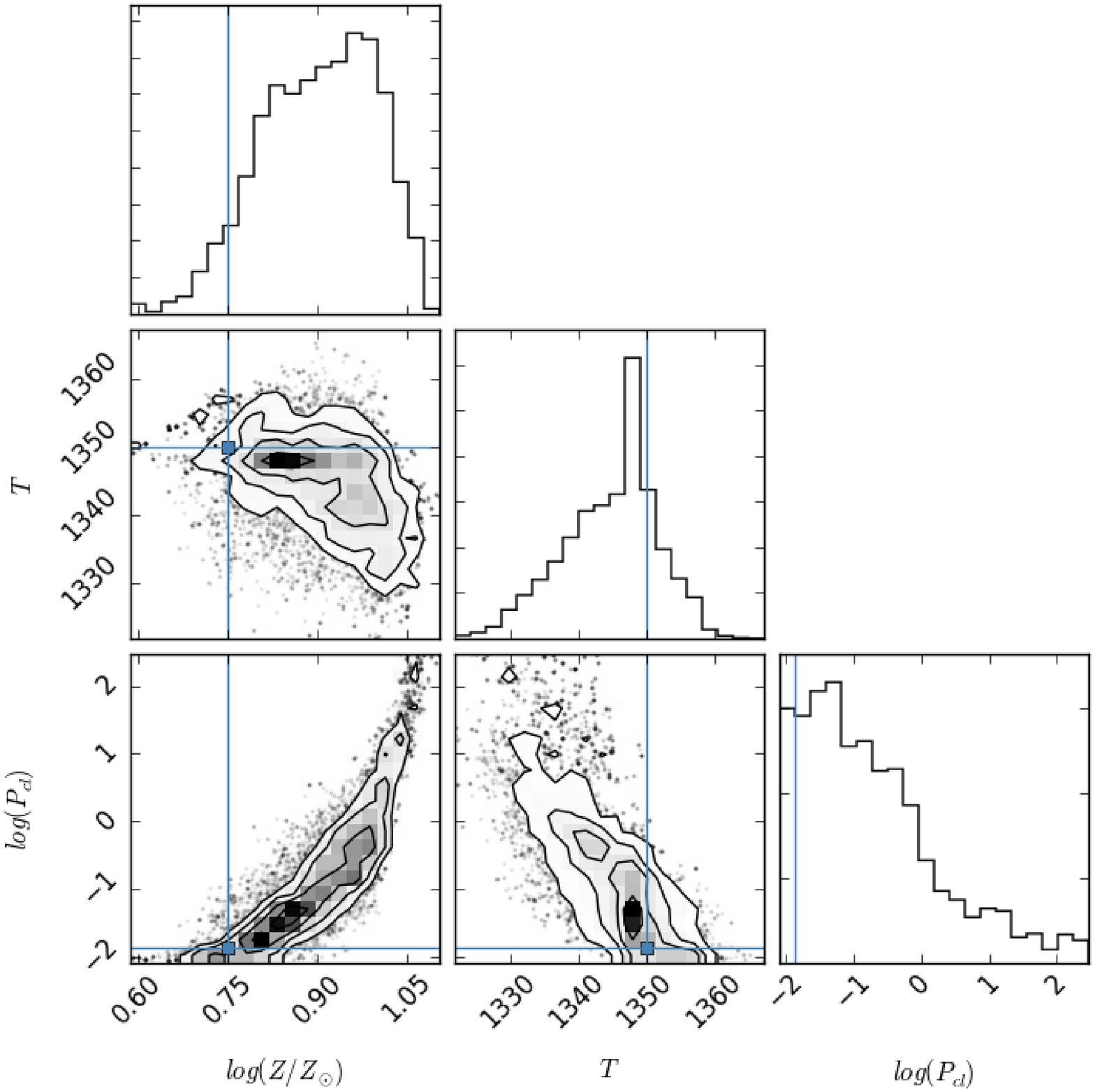}}
\hfill
\subfigure{\includegraphics[width=0.40\textwidth]{prog1.plot.eps}}
\caption{Posterior distributions from retrievals of our three model parameters for selected observations of HAT-P-1b. Each observation uses 10 hours of total observing time. Top left: NIRISS G700XD. Top right: NIRCam F444W. Middle left: NIRSpec G395H. Middle right: MIRI LRS Slitless. Bottom left: MIRI MRS Subchannel A. Bottom right: Program 1.}
\label{ModePlots}
\end{figure*}

Figure \ref{ModePlots} shows posterior distributions for a few representative observations of HAT-P-1b. For the most part, they appear as expected, centered around the correct values of the parameters and narrowing with increasing observing time. However, we note one significant anomaly: the mid-infrared modes are much poorer at recovering the correct parameters, producing much broader posterior distributions, and, in the case of MIRI LRS, multi-modal distributions. This is understandable for LRS because of its lower spectral resolution. However, we do not retrieve either the model parameters or the Poisson noise slope of the mutual information from MRS observations, possibly due to the systematic effects of combining multiple subchannels for the MIRI MRS observing modes.

Table \ref{InfoModes} gives mutual information per degree of freedom estimates for selected individual observing modes representative of the four {\it JWST} instruments and for observing Programs 1$-$5 for the same eight planets. All of these estimates were obtained using the posterior entropy method, since it is more accurate. Interestingly, there are dramatic differences in the mutual information obtained from different modes, but much less so between combinations of modes.

For our simple forward model, the instrument that consistently gives the most information is the NIRISS G700XD mode. This is likely because all of the most sensitive frequency ranges for our three model parameters are in the NIRISS band (see Figure \ref{jacobiansM}). The other modes are not as sensitive and thus do not produce as much information. This means that, at least for simple forward models like this example, NIRISS is one of the most important modes for atmosphere characterization. For this model, NIRSpec is more informative than NIRCam in approximately the same wavelength range and spectral resolution, likely because of its greater sensitivity.

\begin{table*}[htbp]
\caption{Estimates of Mutual Information per Degree of Freedom for Selected Observing Modes}
\begin{center}
\tiny
\begin{tabular}{l|l|r|r|r|r|r|r|l|r|r|r|r|r|r}
\hline
Object    & Mode   & 1.0 h & 1.6 h & 2.5 h & 4.0 h & 6.3 h & 10.0 h & Mode   & 1.0 h & 1.6 h & 2.5 h & 4.0 h & 6.3 h & 10.0 h \\
\hline
HAT-P-12b & G700XD & 3.88 & 4.49 & 4.43 & 4.70 & 5.11 & 5.60 & F444W &      & 2.99 & 3.25 & 3.40 & 3.77 & 4.09 \\
HAT-P-1b  & G700XD & 3.58 & 3.86 & 4.34 & 4.64 & 5.41 & 5.74 & F444W & 1.96 & 1.99 & 2.10 & 2.67 & 2.96 & 3.39 \\
WASP-12b  & G700XD & 3.52 & 3.77 & 4.10 & 4.46 & 4.80 & 5.08 & F444W & 1.33 & 1.31 & 1.63 & 1.73 & 1.87 & 2.77 \\
WASP-17b  & G700XD & 4.87 & 5.18 & 5.45 & 5.86 & 6.15 & 6.47 & F444W & 3.04 & 3.54 & 3.79 & 4.05 & 4.34 & 4.71 \\
WASP-19b  & G700XD & 2.90 & 2.96 & 3.40 & 3.70 & 4.24 & 4.49 & F444W & 1.20 & 1.44 & 1.57 & 1.91 & 1.80 & 2.36 \\
WASP-39b  & G700XD & 4.05 & 4.52 & 4.75 & 5.18 & 5.61 & 5.79 & F444W & 2.47 & 2.43 &      & 3.51 & 3.82 & 4.04 \\
WASP-43b  & G700XD & 2.03 & 2.59 & 3.26 & 2.94 & 4.01 & 3.45 & F444W & 0.82 & 1.03 & 1.11 & 1.48 & 1.66 & 1.86 \\
WASP-6b   & G700XD & 2.62 & 2.80 & 3.22 & 3.65 & 3.91 & 4.25 & F444W & 1.50 &      & 1.47 & 1.89 & 2.32 & 2.45 \\
\hline
HAT-P-12b & G395H  & 2.96 & 3.48 & 3.55 & 3.86 & 4.18 & 4.54 & LRS Slitless & 2.92 & 3.16 & 3.47 & 3.74 & 4.19 & 4.36 \\
HAT-P-1b  & G395H  & 2.20 &      & 2.97 & 3.23 & 3.44 & 3.80 & LRS Slitless &      & 2.12 &      & 2.50 &      & 2.66 \\
WASP-12b  & G395H  & 1.58 & 1.84 & 2.23 & 2.75 & 3.20 & 3.52 & LRS Slitless & 1.64 & 2.19 & 2.69 & 3.04 & 3.35 & 3.73 \\
WASP-17b  & G395H  & 4.23 & 4.50 & 4.87 & 5.17 & 5.51 & 5.86 & LRS Slitless & 4.69 & 4.94 & 5.27 & 5.63 & 5.97 & 6.25 \\
WASP-19b  & G395H  & 1.67 &      & 2.38 & 2.60 & 2.78 & 3.14 & LRS Slitless & 2.01 &      &      & 2.88 &      & 3.37 \\
WASP-39b  & G395H  & 2.96 & 3.27 & 3.56 & 3.89 & 4.30 & 4.56 & LRS Slitless & 3.08 & 3.40 & 3.82 & 4.11 & 4.28 & 4.61 \\
WASP-43b  & G395H  & 1.28 & 1.72 & 1.66 & 1.70 &      & 2.49 & LRS Slitless & 1.13 & 1.44 & 1.17 & 1.74 &      & 1.93 \\
WASP-6b   & G395H  &      & 2.04 &      & 2.57 & 2.85 & 3.15 & LRS Slitless & 2.27 & 2.14 & 2.43 & 2.97 & 3.16 & 3.45 \\
\hline
HAT-P-12b & MIRI A & 5.28 & 5.41 & 5.31 & 5.38 & 5.16 & 5.34 & Prog 1 & 3.70 & 3.78 & 3.95 & 4.40 & 4.69 & 4.87 \\
HAT-P-1b  & MIRI A & 3.64 & 3.76 & 3.79 & 3.79 & 3.35 & 3.95 & Prog 1 & 3.11 & 3.27 & 3.98 & 4.25 & 4.74 & 4.74 \\
WASP-12b  & MIRI A & 4.76 & 4.80 & 4.82 & 4.83 & 4.80 & 4.81 & Prog 1 & 3.09 & 3.16 & 3.52 & 3.83 & 4.16 & 4.51 \\
WASP-17b  & MIRI A & 7.18 & 7.23 & 7.24 & 7.27 & 7.25 & 7.27 & Prog 1 & 4.81 & 5.08 & 5.42 & 5.75 & 6.08 & 6.43 \\
WASP-19b  & MIRI A & 3.18 & 2.98 & 3.28 & 3.01 & 3.38 & 3.40 & Prog 1 & 2.33 & 2.73 & 2.89 & 3.48 & 3.95 & 3.91 \\
WASP-39b  & MIRI A & 5.81 & 5.83 & 5.81 & 5.83 & 5.81 & 5.83 & Prog 1 & 3.81 & 3.81 & 4.20 & 4.49 & 4.89 & 5.25 \\
WASP-43b  & MIRI A & 2.52 & 2.58 &      & 2.70 & 2.54 & 2.47 & Prog 1 & 1.72 &     & 2.24 & 2.47 & 2.79 & -0.33 \\
WASP-6b   & MIRI A & 4.32 & 4.35 & 4.55 & 4.43 & 4.25 & 4.46 & Prog 1 &      & 2.71 & 2.93 & 3.23 & 3.56 & 3.97 \\
\hline
HAT-P-12b & Prog 2 & 3.62 & 4.05 & 4.33 & 4.84 & 4.98 & 5.31 & Prog 3 & 3.11 & 3.48 & 3.72 & 4.19 & 4.26 & 4.53 \\
HAT-P-1b  & Prog 2 & 3.55 & 3.52 & 3.99 & 4.33 & 4.77 & 4.97 & Prog 3 & 2.85 & 3.01 & 3.29 & 3.54 & 4.28 & 4.24 \\
WASP-12b  & Prog 2 & 3.32 & 3.52 & 3.76 & 4.16 & 4.50 & 4.80 & Prog 3 & 2.32 & 2.86 & 3.21 & 3.48 & 3.88 & 4.22 \\
WASP-17b  & Prog 2 & 5.18 & 5.46 & 5.82 & 6.12 & 6.48 & 6.83 & Prog 3 & 4.57 & 4.96 & 5.21 & 5.53 & 5.93 & 6.25 \\
WASP-19b  & Prog 2 & 2.48 & 2.86 & 3.15 & 3.42 & 3.95 & 4.29 & Prog 3 & 1.96 & 2.34 & 2.47 & 2.78 & 3.26 & 3.64 \\
WASP-39b  & Prog 2 & 3.87 & 4.43 & 4.81 & 4.88 & 5.15 & 5.77 & Prog 3 & 3.23 & 3.67 & 3.93 & 4.20 & 4.88 & 4.93 \\
WASP-43b  & Prog 2 & 1.90 & 2.19 & 2.50 & 2.75 & 2.99 & 3.17 & Prog 3 & 1.39 & 1.55 &      & 2.19 & 2.70 & 2.81 \\
WASP-6b   & Prog 2 & 2.70 & 3.03 & 3.20 & 3.62 & 3.72 & 4.13 & Prog 3 & 2.28 & 2.52 & 2.76 &      & 3.19 & 3.47 \\
\hline
HAT-P-12b & Prog 4 & 3.17 & 4.13 & 3.92 & 4.21 & 4.59 & 4.83 & Prog 5 & 3.34 & 3.64 & 3.84 & 4.56 & 4.60 & 4.91 \\
HAT-P-1b  & Prog 4 & 2.98 & 3.69 & 3.57 & 3.79 & 4.03 & 4.69 & Prog 5 & 2.94 & 3.19 & 3.33 & 3.73 & 4.10 & 4.48 \\
WASP-12b  & Prog 4 & 2.79 & 3.03 & 3.37 & 3.72 & 4.09 & 4.41 & Prog 5 & 2.71 & 3.34 & 3.28 & 3.70 & 4.03 & 4.41 \\
WASP-17b  & Prog 4 & 4.69 & 5.06 & 5.47 & 5.75 & 6.11 & 6.44 & Prog 5 & 4.73 & 5.05 & 5.36 & 5.71 & 6.00 & 6.38 \\
WASP-19b  & Prog 4 & 2.17 & 2.63 & 2.94 & 3.23 & 3.53 & 3.78 & Prog 5 & 2.10 & 2.63 & 2.65 & 2.96 & 3.20 & 3.47 \\
WASP-39b  & Prog 4 & 3.51 & 3.78 & 4.34 & 4.68 & 4.82 & 5.21 & Prog 5 & 3.46 & 3.96 & 4.17 & 4.68 & 4.81 & 4.97 \\
WASP-43b  & Prog 4 & 1.77 & 1.72 & 2.21 &      & 2.54 & 3.06 & Prog 5 & 1.82 & 1.95 & 2.01 & 2.46 & 2.49 & 3.04 \\
WASP-6b   & Prog 4 & 2.37 & 2.56 & 2.86 & 3.16 & 3.33 & 3.63 & Prog 5 & 2.19 & 2.62 & 3.46 & 3.08 & 3.25 & 3.67 \\
\hline
\end{tabular}
\end{center}
\label{InfoModes}
\end{table*}

In comparing the numbered observing programs, we must consider that NIRISS is the most informative single mode for this model set in Programs 1 and 2 by a significant margin. Program 2 uses only three visits compared with four for the same total observing time for Program 1, so more observing time is devoted to NIRISS in Program 2. For our particular model, NIRISS provides such an advantage over the other modes that giving it more observing time outweighs the loss of coverage in the 2.5-2.9 $\mu m$ range from Program 1 and provides slightly more mutual information overall based on the longer observing time on NIRISS.

Programs 3 and 5 both rely on NIRSpec rather than NIRCam. The important wavelength range of 4-5 $\mu m$ is still covered, and NIRSpec is more informative than NIRCam, so we might expect these programs to provide more information for our simple forward model set than Programs 1 and 2. However, Programs 3 and 5 require 7 and 9 visits, respectively, so they allow less observing time for the short-wavelength observations that are most informative for our model, and this again proves to be the dominant factor, making Programs 3 and 5 less informative than Programs 1 and 2.

Program 4 does not use NIRISS at all, as with the similar Program 3. However, it requires fewer visits (5) than Program 3, allowing more observing time for short-wavelength observations, so that it performs better than Program 3. Interestingly, it also performs slightly better than Program 5, which does use NIRISS, possibly because it allows more observing time for the high-resolution G395H mode.

Taken together, these results suggest that if a single observing mode is known to be more informative than the others by a significant margin ($\sim 1$ bit per degree of freedom in this case), it is better to observe with that mode alone than to use several modes, for a given total observing time, because shortening the observing time for the best mode reduces the statistics. However, we also note that 2.5 hours of observation with Program 4, which does not use NIRISS, the most informative mode for our model set, and which only observes half an hour per mode, still provides more mutual information than 1 hour of observation with NIRISS alone. In other words, for our forward model set, there is more to be gained by increasing the number of observing modes and the observing time together than there is by increasing either of them separately, a property that will be worth exploring in the context of other forward models.

\begin{figure}[htp]
\includegraphics[width=\columnwidth]{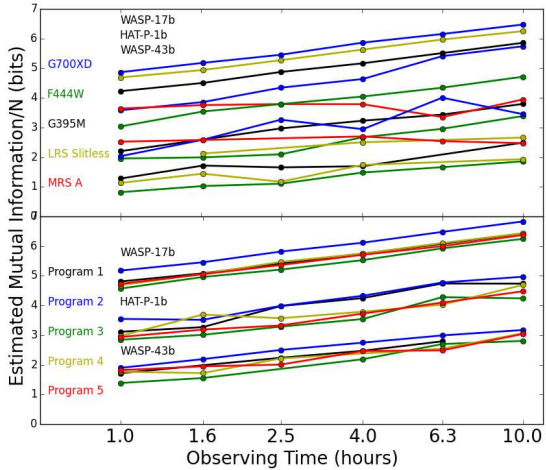}
\caption{Estimated mutual information per degree of freedom based on posterior entropy versus observing time for selected {\it JWST} observing modes representive of all four spectroscopic instruments and our numbered programs. Results for WASP-17b, HAT-P-1b, and WASP-43b are plotted, which yield the largest, smallest, and near-median amounts of mutual information, respectively, out of the planets we study.}
\label{InfoSingleFig}
\end{figure}

Figure \ref{InfoSingleFig} shows the estimated mutual information per degree of freedom versus observing time for each of these observing programs for three selected objects: WASP-17b, WASP-43b, and HAT-P-1b, which yield the largest, smallest, and near-median amounts of mutual information per degree of freedom, respectively. This illustrates more clearly that all of our numbered observing programs yield about the same amount of mutual information per degree of freedom when compared with the individual modes, which show much more variation. This is unsurprising given that, for our simple forward model, the amount of mutual information obtained is strongly dependent on the most informative mode in the observing program.

These results are applicable to our very simple atmosphere model with only three parameters, and other model sets will produce different results that may require greater finesse to optimize an observing program to measure them. In particular, different molecular species may have absorption bands in the wavelength ranges of different modes, so a different combination of observing modes and instruments would be necessary to measure their chemical abundances. However, these same methods will apply for determining the best modalities for observing them.

\subsection{Bright Objects: HD 189733b and HD 209458b}
\label{bright}

\begin{figure}[htp]
\includegraphics[width=\columnwidth]{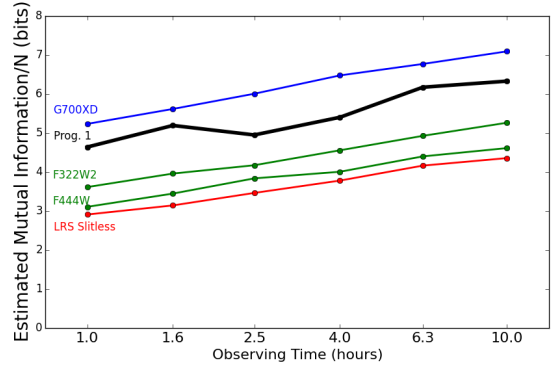}
\caption{Estimated mutual information per degree of freedom based on posterior entropy versus observing time for our Program 1 and its constituent observing modes for HD 209458b.}
\label{InfoBrightFig}
\end{figure}

HD 189733 and HD 209458 fall into the category of bright targets ($J<8$) that are too bright to be observed with NIRSpec, which means that only Program 1 from our numbered programs can be used for them. We list the estimated mutual information per degree of freedom for Program 1 observations of HD 209458b and the constituent observing modes in Table \ref{InfoBright} and plot them in Figure \ref{InfoBrightFig}. Predictably, the mutual information obtained with observations of this object is significantly greater than for fainter objects. More notably, the observations still do not reach a point of diminishing returns after 10 hours. This is likely to change, however, for more complex model sets with more degrees of freedom.

\begin{table*}[htbp]
\caption{Estimated Mutual Information per Degree of Freedom for Observations of HD 209458b}
\begin{center}
\begin{tabular}{l|l|r|r|r|r|r|r}
\hline
Mode      & Estimate   & 1.0 h & 1.6 h & 2.5 h & 4.0 h & 6.3 h & 10.0 h \\
\hline
Program 1 & Covariance &  4.71 &  5.32 &  5.05 &  5.58 &  6.32 &  6.50 \\
Program 1 & Entropy    &  4.64 &  5.19 &  4.95 &  5.40 &  6.17 &  6.33 \\
G700XD    & Covariance &  5.34 &  5.74 &  6.17 &  6.65 &  6.96 &  7.28 \\
G700XD    & Entropy    &  5.23 &  5.62 &  6.01 &  6.48 &  6.77 &  7.09 \\
F322W2    & Covariance &  3.83 &  3.90 &  4.37 &  4.67 &  4.95 &  5.38 \\
F322W2    & Entropy    &  3.61 &  3.96 &  4.17 &  4.56 &  4.93 &  5.26 \\
F444W     & Covariance &  3.32 &  3.59 &  3.64 &  4.10 &  4.43 &  4.81 \\
F444W     & Entropy    &  3.11 &  3.45 &  3.84 &  4.01 &  4.40 &  4.61 \\
MIRI LRS  & Covariance &  2.70 &  3.45 &  3.17 &  3.76 &  4.12 &  4.45 \\
MIRI LRS  & Entropy    &  2.91 &  3.14 &  3.46 &  3.78 &  4.16 &  4.36 \\
\hline
\end{tabular}
\end{center}
\label{InfoBright}
\end{table*}

\subsection{Faint Objects: Kepler-7b}
\label{faint}

\begin{table*}[htbp]
\caption{Estimated Mutual Information per Degree of Freedom for Observations of Kepler-7b}
\begin{center}
\begin{tabular}{l|l|r|r|r|r|r|r}
\hline
Mode      & Estimate   & 1.0 h & 1.6 h & 2.5 h & 4.0 h & 6.3 h & 10.0 h \\
\hline
Program 6 & Covariance &  1.83 &  2.10 &  2.04 &  2.63 &  2.90 &  3.31 \\
Program 6 & Entropy    &  1.76 &  1.98 &  2.21 &  2.57 &  2.81 &  3.17 \\
Prism (Prog. 7) & Covariance   &  2.40 &  2.32 &  2.85 &  2.85 &  3.51 &  3.83 \\
Prism (Prog. 7) & Entropy      &  2.32 &  2.28 &  2.76 &  2.78 &  3.31 &  3.64 \\
MIRI LRS  & Covariance &  1.61 &  2.07 &       &  2.08 &  2.27 &  2.96 \\
MIRI LRS  & Entropy    &  1.75 &  1.99 &       &  2.28 &  2.58 &  3.17 \\
\hline
\end{tabular}
\end{center}
\label{InfoFaint}
\end{table*}

Stars fainter than $J\sim 11$ can be best characterized spectroscopically with the NIRSpec Prism and MIRI LRS which have lower spectral resolution and thus more light collection power per resolution unit. Planets with host stars in this brightness range are not well studied. However, we can still model the retrievals with these two modes. We show the results of these retrievals in Table \ref{InfoFaint}. Here, we find that the Prism mode is significantly more informative than LRS, probably because it covers the important short wavelengths. As a result, it is more efficient to observe with only the Prism (our Program 7) than with both modes (Program 6). In this case, this is true even if the total observing time is increased to compensate: one hour each with the Prism and MIRI LRS is not an advantage over one hour with the Prism alone. Once again, we do not see diminishing returns after 10 hours.

% % % % % % % % % % % % % % % % % % % % % % % % % % % % % % % % % % % % % % % % % % % % % % % % % % % % % % % % % % % % % % % % % % % % % % %

\section{Conclusions}
\label{summary}

We have presented a methodology to design optimized observing programs for transiting planets with {\it JWST} and other observing platforms using retrievals of model parameters and a novel statistical analysis. We can use the idea of mutual information, or more precisely, mutual information per degree of freedom, to get a sense of how to optimize observing modalities with {\it JWST}. The mutual information per degree of freedom is closely related to the constraints placed on the atmosphere model parameters by a given observation.

We have investigated a range of modalities for observing transiting planets with {\it JWST} in order to characterize their atmospheres. To explore the capabilities of {\it JWST}, we employed a simple three-parameter atmosphere model with parameters for metallicity, isothermal temperature, and cloud top pressure. In practice, forward models will be more complex than this, but the methods will be the same, and this study reveals some clear trends and strategies.

Our statistical analysis was done with an MCMC package, which found the posterior distribution of the model parameters for a given observation. We introduce the use of the mutual information between the prior and posterior distributions per degree of freedom as the figure of merit for optimization.

%We estimate the mutual information using two methods, and we find that the method of computing the posterior entropy for a single observation is more accurate than using the covariance matrix.

We estimated the mutual information for all of the applicable observing modes and several observing programs for eleven hot jupiters and analyzed the results, comparing observing modalities with the same cost in terms of integration time to find the most efficient one. Our results are specific to our simple forward model set, and other models will produce different results. However, while we cannot advocate a paricular instrument or observing program, we present this methodology as the most rigorous way to optimize an observing program for a given atmosphere model.

To examine specific modalities for observation, we performed retrievals of our three model parameters with all of the individual observing modes and seven observing programs constituting different combinations of modes. We found that the Slitless mode of MIRI LRS provides more mutual information than the Slit mode, as would be expected given its greater light collection power, so we used the Slitless mode in all applicable observing programs. We also found that increasing the integration time continues to improve the statistics of our results according to our estimate of Poisson noise without systematic effects for reasonable JWST observing times.

In selecting specific modes for observation, we find that the method that yields the greatest information gain in terms of quantitative mutual information is to observe in the single mode that is the most informative per unit time for a given model. For our simple, three-parameter model set, NIRISS G700XD consistently provides the most mutual information for a given integration time compared with other observing modes and program. However, given our simple forward model, we caution that there may be an advantage in selecting observing modes to characterize more parameters for more complex models, such as those that include chemical abundances.

%[Flag!] Qualify that this is for our simple model set, and other models will be different. Say from the beginning that we are not advocating a particular instrument. We are advocating the method. It's the modality of trying to choose how to optimize observing programs that we are advocating. Recraft the entire conclusion to reflect this.

%In contrast, it is significantly less efficient to use more observing modes to achieve wider spectral coverage rather than selecting the smallest number of modes that well characterize the model parameters.

Other than their direct role in determining the pressure scale height, the atmospheric temperature and gravity have little direct additional effect on the fits. They can affect the amount of information obtained from an observing program, but they do not change the best overall modalities for observing. The same is true of the ``cloudiness'' of the atmosphere, with cloudier atmospheres reducing the amount of information that can be retrieved, but not changing the optimal observing modalities. We also study three distinct ranges for target brightness, $J<8$, $8<J<11$, and $J>11$, all of which require different sets of observing modes, and we find the same trends in each range within the set of available observing modes. However, with only two significant modes in the $J>11$ range, the range of possible observing programs is more limited. For our model set, the NIRSpec Prism  alone is the best method to characterize faint objects.

More complex forward models than our three-parameter example will result in different numbers for the mutual information provided by each observation, but the overall strategy should remain the same. Designing an observing program according to these methods and trends will make better use of {\it JWST} observing time for characterizing exoplanet atmospheres.

% % % % % % % % % % % % % % % % % % % % % % % % % % % % % % % % % % % % % % % % % % % % % % % % % % % % % % % % % % % % % % % % % % % % % % %

\acknowledgements

The authors would like to acknowledge support from NASA Grant NNX15AE19G; NASA, WFIRST-SIT Sponsor Award \#61238715-122362 (Prime \# NASA NNG16PJ24C); and NASA, JPL subcontract nos. 1538907 \& 1529729. We thank Tom Greene and Jonathan Fraine for providing throughput values for the NIRCam grisms. We also thank Tom Greene along with Lo\"{i}c Albert and Everett Schlawin for providing corrections to throughput values and saturation limits.

% % % % % % % % % % % % % % % % % % % % % % % % % % % % % % % % % % % % % % % % % % % % % % % % % % % % % % % % % % % % % % % % % % % % % % %

%\clearpage
\bibliographystyle{apj}
\bibliography{apj-jour,refs}

\begin{thebibliography}{69}
\expandafter\ifx\csname natexlab\endcsname\relax\def\natexlab#1{#1}\fi

\bibitem[{{Anderson} {et~al.}(2010){Anderson}, {Hellier}, {Gillon}, {Triaud},
  {Smalley}, {Hebb}, {Collier Cameron}, {Maxted}, {Queloz}, {West}, {Bentley},
  {Enoch}, {Horne}, {Lister}, {Mayor}, {Parley}, {Pepe}, {Pollacco},
  {S{\'e}gransan}, {Udry}, \& {Wilson}}]{2010ApJ...709..159A}
{Anderson}, D.~R., {et~al.} 2010, \apj, 709, 159

\bibitem[{{Bakos} {et~al.}(2004){Bakos}, {Noyes}, {Kov{\'a}cs}, {Stanek},
  {Sasselov}, \& {Domsa}}]{2004PASP..116..266B}
{Bakos}, G., {Noyes}, R.~W., {Kov{\'a}cs}, G., {Stanek}, K.~Z., {Sasselov},
  D.~D., \& {Domsa}, I. 2004, \pasp, 116, 266

\bibitem[{{Bakos} {et~al.}(2007){Bakos}, {Noyes}, {Kov{\'a}cs}, {Latham},
  {Sasselov}, {Torres}, {Fischer}, {Stefanik}, {Sato}, {Johnson}, {P{\'a}l},
  {Marcy}, {Butler}, {Esquerdo}, {Stanek}, {L{\'a}z{\'a}r}, {Papp}, {S{\'a}ri},
  \& {Sip{\H o}cz}}]{2007ApJ...656..552B}
{Bakos}, G.~{\'A}., {et~al.} 2007, \apj, 656, 552

\bibitem[{{Barstow} {et~al.}(2015){Barstow}, {Aigrain}, {Irwin}, {Kendrew}, \&
  {Fletcher}}]{2015MNRAS.448.2546B}
{Barstow}, J.~K., {Aigrain}, S., {Irwin}, P.~G.~J., {Kendrew}, S., \&
  {Fletcher}, L.~N. 2015, \mnras, 448, 2546

\bibitem[{{Beaulieu} {et~al.}(2010){Beaulieu}, {Kipping}, {Batista}, {Tinetti},
  {Ribas}, {Carey}, {Noriega-Crespo}, {Griffith}, {Campanella}, {Dong},
  {Tennyson}, {Barber}, {Deroo}, {Fossey}, {Liang}, {Swain}, {Yung}, \&
  {Allard}}]{2010MNRAS.409..963B}
{Beaulieu}, J.~P., {et~al.} 2010, \mnras, 409, 963

\bibitem[{{Beichman} {et~al.}(2014){Beichman}, {Benneke}, {Knutson}, {Smith},
  {Lagage}, {Dressing}, {Latham}, {Lunine}, {Birkmann}, {Ferruit}, {Giardino},
  {Kempton}, {Carey}, {Krick}, {Deroo}, {Mandell}, {Ressler}, {Shporer},
  {Swain}, {Vasisht}, {Ricker}, {Bouwman}, {Crossfield}, {Greene}, {Howell},
  {Christiansen}, {Ciardi}, {Clampin}, {Greenhouse}, {Sozzetti}, {Goudfrooij},
  {Hines}, {Keyes}, {Lee}, {McCullough}, {Robberto}, {Stansberry}, {Valenti},
  {Rieke}, {Rieke}, {Fortney}, {Bean}, {Kreidberg}, {Ehrenreich}, {Deming},
  {Albert}, {Doyon}, \& {Sing}}]{2014PASP..126.1134B}
{Beichman}, C., {et~al.} 2014, \pasp, 126, 1134

\bibitem[{{Benneke}(2015)}]{2015arXiv150407655B}
{Benneke}, B. 2015, ArXiv e-prints

\bibitem[{{Birkmann} {et~al.}(2014){Birkmann}, {Ferruit}, {Alves de Oliveira},
  {B{\"o}ker}, {De Marchi}, {Giardino}, {Sirianni}, {Stuhlinger}, {Jensen},
  {Rumler}, {Falcolini}, {te Plate}, {Cresci}, {Dorner}, {Ehrenwinkler},
  {Gnata}, \& {Wettemann}}]{2014SPIE.9143E..08B}
{Birkmann}, S.~M., {et~al.} 2014, in \procspie, Vol. 9143, Space Telescopes and
  Instrumentation 2014: Optical, Infrared, and Millimeter Wave, 914308

\bibitem[{{Bouchy} {et~al.}(2005){Bouchy}, {Udry}, {Mayor}, {Moutou}, {Pont},
  {Iribarne}, {da Silva}, {Ilovaisky}, {Queloz}, {Santos}, {S{\'e}gransan}, \&
  {Zucker}}]{2005A&A...444L..15B}
{Bouchy}, F., {et~al.} 2005, \aap, 444, L15

\bibitem[{{Burrows} {et~al.}(2007){Burrows}, {Hubeny}, {Budaj}, {Knutson}, \&
  {Charbonneau}}]{2007ApJ...668L.171B}
{Burrows}, A., {Hubeny}, I., {Budaj}, J., {Knutson}, H.~A., \& {Charbonneau},
  D. 2007, \apjl, 668, L171

\bibitem[{{Burrows}(2014)}]{2014PNAS..11112601B}
{Burrows}, A.~S. 2014, Proceedings of the National Academy of Science, 111,
  12601

\bibitem[{{Castellano} {et~al.}(2000){Castellano}, {Jenkins}, {Trilling},
  {Doyle}, \& {Koch}}]{2000ApJ...532L..51C}
{Castellano}, T., {Jenkins}, J., {Trilling}, D.~E., {Doyle}, L., \& {Koch}, D.
  2000, \apjl, 532, L51

\bibitem[{Chaloner \& Verdinelli(1995)}]{Chaloner95bayesianexperimental}
Chaloner, K., \& Verdinelli, I. 1995, Statistical Science, 10, 273

\bibitem[{{Charbonneau} {et~al.}(2002){Charbonneau}, {Brown}, {Noyes}, \&
  {Gilliland}}]{2002ApJ...568..377C}
{Charbonneau}, D., {Brown}, T.~M., {Noyes}, R.~W., \& {Gilliland}, R.~L. 2002,
  \apj, 568, 377

\bibitem[{{Coughlin} {et~al.}(2016){Coughlin}, {Mullally}, {Thompson}, {Rowe},
  {Burke}, {Latham}, {Batalha}, {Ofir}, {Quarles}, {Henze}, {Wolfgang},
  {Caldwell}, {Bryson}, {Shporer}, {Catanzarite}, {Akeson}, {Barclay},
  {Borucki}, {Boyajian}, {Campbell}, {Christiansen}, {Girouard}, {Haas},
  {Howell}, {Huber}, {Jenkins}, {Li}, {Patil-Sabale}, {Quintana}, {Ramirez},
  {Seader}, {Smith}, {Tenenbaum}, {Twicken}, \&
  {Zamudio}}]{2016ApJS..224...12C}
{Coughlin}, J.~L., {et~al.} 2016, \apjs, 224, 12

\bibitem[{{Danielski} {et~al.}(2014){Danielski}, {Deroo}, {Waldmann}, {Hollis},
  {Tinetti}, \& {Swain}}]{2014ApJ...785...35D}
{Danielski}, C., {Deroo}, P., {Waldmann}, I.~P., {Hollis}, M.~D.~J., {Tinetti},
  G., \& {Swain}, M.~R. 2014, \apj, 785, 35

\bibitem[{{Deming} {et~al.}(2009){Deming}, {Seager}, {Winn}, {Miller-Ricci},
  {Clampin}, {Lindler}, {Greene}, {Charbonneau}, {Laughlin}, {Ricker},
  {Latham}, \& {Ennico}}]{2009PASP..121..952D}
{Deming}, D., {et~al.} 2009, \pasp, 121, 952

\bibitem[{{Deming} {et~al.}(2013){Deming}, {Wilkins}, {McCullough}, {Burrows},
  {Fortney}, {Agol}, {Dobbs-Dixon}, {Madhusudhan}, {Crouzet}, {Desert},
  {Gilliland}, {Haynes}, {Knutson}, {Line}, {Magic}, {Mandell}, {Ranjan},
  {Charbonneau}, {Clampin}, {Seager}, \& {Showman}}]{2013ApJ...774...95D}
---. 2013, \apj, 774, 95

\bibitem[{{D{\'e}sert} {et~al.}(2009){D{\'e}sert}, {Lecavelier des Etangs},
  {H{\'e}brard}, {Sing}, {Ehrenreich}, {Ferlet}, \&
  {Vidal-Madjar}}]{2009ApJ...699..478D}
{D{\'e}sert}, J.-M., {Lecavelier des Etangs}, A., {H{\'e}brard}, G., {Sing},
  D.~K., {Ehrenreich}, D., {Ferlet}, R., \& {Vidal-Madjar}, A. 2009, \apj, 699,
  478

\bibitem[{{Doyon} {et~al.}(2012){Doyon}, {Hutchings}, {Beaulieu}, {Albert},
  {Lafreni{\`e}re}, {Willott}, {Touahri}, {Rowlands}, {Maszkiewicz},
  {Fullerton}, {Volk}, {Martel}, {Chayer}, {Sivaramakrishnan}, {Abraham},
  {Ferrarese}, {Jayawardhana}, {Johnstone}, {Meyer}, {Pipher}, \&
  {Sawicki}}]{2012SPIE.8442E..2RD}
{Doyon}, R., {et~al.} 2012, in \procspie, Vol. 8442, Space Telescopes and
  Instrumentation 2012: Optical, Infrared, and Millimeter Wave, 84422R

\bibitem[{{Faedi} {et~al.}(2011){Faedi}, {Barros}, {Anderson}, {Brown},
  {Collier Cameron}, {Pollacco}, {Boisse}, {H{\'e}brard}, {Lendl}, {Lister},
  {Smalley}, {Street}, {Triaud}, {Bento}, {Bouchy}, {Butters}, {Enoch},
  {Haswell}, {Hellier}, {Keenan}, {Miller}, {Moulds}, {Moutou}, {Norton},
  {Queloz}, {Santerne}, {Simpson}, {Skillen}, {Smith}, {Udry}, {Watson},
  {West}, \& {Wheatley}}]{2011A&A...531A..40F}
{Faedi}, F., {et~al.} 2011, \aap, 531, A40

\bibitem[{{Foreman-Mackey} {et~al.}(2013){Foreman-Mackey}, {Hogg}, {Lang}, \&
  {Goodman}}]{2013PASP..125..306F}
{Foreman-Mackey}, D., {Hogg}, D.~W., {Lang}, D., \& {Goodman}, J. 2013, \pasp,
  125, 306

\bibitem[{{Gillon} {et~al.}(2009){Gillon}, {Anderson}, {Triaud}, {Hellier},
  {Maxted}, {Pollaco}, {Queloz}, {Smalley}, {West}, {Wilson}, {Bentley},
  {Collier Cameron}, {Enoch}, {Hebb}, {Horne}, {Irwin}, {Joshi}, {Lister},
  {Mayor}, {Pepe}, {Parley}, {Segransan}, {Udry}, \&
  {Wheatley}}]{2009A&A...501..785G}
{Gillon}, M., {et~al.} 2009, \aap, 501, 785

\bibitem[{{Glasse} {et~al.}(2015){Glasse}, {Rieke}, {Bauwens},
  {Garc{\'{\i}}a-Mar{\'{\i}}n}, {Ressler}, {Rost}, {Tikkanen}, {Vandenbussche},
  \& {Wright}}]{2015PASP..127..686G}
{Glasse}, A., {et~al.} 2015, \pasp, 127, 686

\bibitem[{{Greene} {et~al.}(2007){Greene}, {Beichman}, {Eisenstein}, {Horner},
  {Kelly}, {Mao}, {Meyer}, {Rieke}, \& {Shi}}]{2007SPIE.6693E..0GG}
{Greene}, T., {et~al.} 2007, in \procspie, Vol. 6693, Techniques and
  Instrumentation for Detection of Exoplanets III, 66930G

\bibitem[{{Greene} {et~al.}(2016){Greene}, {Line}, {Montero}, {Fortney},
  {Lustig-Yaeger}, \& {Luther}}]{2016ApJ...817...17G}
{Greene}, T.~P., {Line}, M.~R., {Montero}, C., {Fortney}, J.~J.,
  {Lustig-Yaeger}, J., \& {Luther}, K. 2016, \apj, 817, 17

\bibitem[{{Hansen} {et~al.}(2014){Hansen}, {Schwartz}, \&
  {Cowan}}]{2014MNRAS.444.3632H}
{Hansen}, C.~J., {Schwartz}, J.~C., \& {Cowan}, N.~B. 2014, \mnras, 444, 3632

\bibitem[{{Hartman} {et~al.}(2009){Hartman}, {Bakos}, {Torres}, {Kov{\'a}cs},
  {Noyes}, {P{\'a}l}, {Latham}, {Sip{\H o}cz}, {Fischer}, {Johnson}, {Marcy},
  {Butler}, {Howard}, {Esquerdo}, {Sasselov}, {Kov{\'a}cs}, {Stefanik},
  {Fernandez}, {L{\'a}z{\'a}r}, {Papp}, \& {S{\'a}ri}}]{2009ApJ...706..785H}
{Hartman}, J.~D., {et~al.} 2009, \apj, 706, 785

\bibitem[{{Hebb} {et~al.}(2009){Hebb}, {Collier-Cameron}, {Loeillet},
  {Pollacco}, {H{\'e}brard}, {Street}, {Bouchy}, {Stempels}, {Moutou},
  {Simpson}, {Udry}, {Joshi}, {West}, {Skillen}, {Wilson}, {McDonald},
  {Gibson}, {Aigrain}, {Anderson}, {Benn}, {Christian}, {Enoch}, {Haswell},
  {Hellier}, {Horne}, {Irwin}, {Lister}, {Maxted}, {Mayor}, {Norton}, {Parley},
  {Pont}, {Queloz}, {Smalley}, \& {Wheatley}}]{2009ApJ...693.1920H}
{Hebb}, L., {et~al.} 2009, \apj, 693, 1920

\bibitem[{{Hebb} {et~al.}(2010){Hebb}, {Collier-Cameron}, {Triaud}, {Lister},
  {Smalley}, {Maxted}, {Hellier}, {Anderson}, {Pollacco}, {Gillon}, {Queloz},
  {West}, {Bentley}, {Enoch}, {Haswell}, {Horne}, {Mayor}, {Pepe}, {Segransan},
  {Skillen}, {Udry}, \& {Wheatley}}]{2010ApJ...708..224H}
---. 2010, \apj, 708, 224

\bibitem[{{Hellier} {et~al.}(2011){Hellier}, {Anderson}, {Collier Cameron},
  {Gillon}, {Jehin}, {Lendl}, {Maxted}, {Pepe}, {Pollacco}, {Queloz},
  {S{\'e}gransan}, {Smalley}, {Smith}, {Southworth}, {Triaud}, {Udry}, \&
  {West}}]{2011A&A...535L...7H}
{Hellier}, C., {et~al.} 2011, \aap, 535, L7

\bibitem[{{Howe} \& {Burrows}(2012)}]{2012ApJ...756..176H}
{Howe}, A.~R., \& {Burrows}, A.~S. 2012, \apj, 756, 176

\bibitem[{{Irwin} {et~al.}(2008){Irwin}, {Teanby}, {de Kok}, {Fletcher},
  {Howett}, {Tsang}, {Wilson}, {Calcutt}, {Nixon}, \&
  {Parrish}}]{2008JQSRT.109.1136I}
{Irwin}, P.~G.~J., {et~al.} 2008, \jqsrt, 109, 1136

\bibitem[{{Kendrew} {et~al.}(2015){Kendrew}, {Scheithauer}, {Bouchet},
  {Amiaux}, {Azzollini}, {Bouwman}, {Chen}, {Dubreuil}, {Fischer}, {Glasse},
  {Greene}, {Lagage}, {Lahuis}, {Ronayette}, {Wright}, \&
  {Wright}}]{2015PASP..127..623K}
{Kendrew}, S., {et~al.} 2015, \pasp, 127, 623

\bibitem[{{Knutson}(2007)}]{2007Natur.448..143K}
{Knutson}, H.~A. 2007, \nat, 448, 143

\bibitem[{Kullback \& Leibler(1951)}]{Kullback51klDivergence}
Kullback, S., \& Leibler, R.~A. 1951, Ann. Math. Statist., 22, 79

\bibitem[{{Latham} {et~al.}(2010){Latham}, {Borucki}, {Koch}, {Brown},
  {Buchhave}, {Basri}, {Batalha}, {Caldwell}, {Cochran}, {Dunham}, {F{\H
  u}r{\'e}sz}, {Gautier}, {Geary}, {Gilliland}, {Howell}, {Jenkins},
  {Lissauer}, {Marcy}, {Monet}, {Rowe}, \& {Sasselov}}]{2010ApJ...713L.140L}
{Latham}, D.~W., {et~al.} 2010, \apjl, 713, L140

\bibitem[{Liepe {et~al.}(2013)Liepe, Filippi, Komorowski, \&
  Stumpf}]{journals/ploscb/LiepeFKS13}
Liepe, J., Filippi, S., Komorowski, M., \& Stumpf, M. P.~H. 2013, PLoS
  Computational Biology, 9

\bibitem[{Lindley(1956)}]{lindley1956}
Lindley, D.~V. 1956, Annals of Mathematical Statistics, 27, 986

\bibitem[{{Line} \& {Parmentier}(2016)}]{2016ApJ...820...78L}
{Line}, M.~R., \& {Parmentier}, V. 2016, \apj, 820, 78

\bibitem[{{Line} \& {Yung}(2013)}]{2013ApJ...779....3L}
{Line}, M.~R., \& {Yung}, Y.~L. 2013, \apj, 779, 3

\bibitem[{{Line} {et~al.}(2012){Line}, {Zhang}, {Vasisht}, {Natraj}, {Chen}, \&
  {Yung}}]{2012ApJ...749...93L}
{Line}, M.~R., {Zhang}, X., {Vasisht}, G., {Natraj}, V., {Chen}, P., \& {Yung},
  Y.~L. 2012, \apj, 749, 93

\bibitem[{{Line} {et~al.}(2013){Line}, {Wolf}, {Zhang}, {Knutson}, {Kammer},
  {Ellison}, {Deroo}, {Crisp}, \& {Yung}}]{2013ApJ...775..137L}
{Line}, M.~R., {et~al.} 2013, \apj, 775, 137

\bibitem[{{Mandell} {et~al.}(2013){Mandell}, {Haynes}, {Sinukoff},
  {Madhusudhan}, {Burrows}, \& {Deming}}]{2013ApJ...779..128M}
{Mandell}, A.~M., {Haynes}, K., {Sinukoff}, E., {Madhusudhan}, N., {Burrows},
  A., \& {Deming}, D. 2013, \apj, 779, 128

\bibitem[{{Maurin} {et~al.}(2012){Maurin}, {Selsis}, {Hersant}, \&
  {Belu}}]{2012A&A...538A..95M}
{Maurin}, A.~S., {Selsis}, F., {Hersant}, F., \& {Belu}, A. 2012, \aap, 538,
  A95

\bibitem[{{McCullough} {et~al.}(2014){McCullough}, {Crouzet}, {Deming}, \&
  {Madhusudhan}}]{2014ApJ...791...55M}
{McCullough}, P.~R., {Crouzet}, N., {Deming}, D., \& {Madhusudhan}, N. 2014,
  \apj, 791, 55

\bibitem[{{Pollacco} {et~al.}(2006){Pollacco}, {Skillen}, {Collier Cameron},
  {Christian}, {Hellier}, {Irwin}, {Lister}, {Street}, {West}, {Anderson},
  {Clarkson}, {Deeg}, {Enoch}, {Evans}, {Fitzsimmons}, {Haswell}, {Hodgkin},
  {Horne}, {Kane}, {Keenan}, {Maxted}, {Norton}, {Osborne}, {Parley}, {Ryans},
  {Smalley}, {Wheatley}, \& {Wilson}}]{2006PASP..118.1407P}
{Pollacco}, D.~L., {et~al.} 2006, \pasp, 118, 1407

\bibitem[{{Pont} {et~al.}(2013){Pont}, {Sing}, {Gibson}, {Aigrain}, {Henry}, \&
  {Husnoo}}]{2013MNRAS.432.2917P}
{Pont}, F., {Sing}, D.~K., {Gibson}, N.~P., {Aigrain}, S., {Henry}, G., \&
  {Husnoo}, N. 2013, \mnras, 432, 2917

\bibitem[{{Richardson} {et~al.}(2006){Richardson}, {Harrington}, {Seager}, \&
  {Deming}}]{2006ApJ...649.1043R}
{Richardson}, L.~J., {Harrington}, J., {Seager}, S., \& {Deming}, D. 2006,
  \apj, 649, 1043

\bibitem[{{Rodgers}(2000)}]{2000SAOPP...2.....R}
{Rodgers}, C.~D. 2000, Inverse Methods for Atmospheric Sounding - Theory and
  Practice.~Series: Series on Atmospheric Oceanic and Planetary Physics, ISBN:
  <ISBN>9789812813718</ISBN>.~World Scientific Publishing Co.~Pte.~Ltd., Edited
  by Clive D.~Rodgers, vol.~2, 2

\bibitem[{{Samuel} {et~al.}(2014){Samuel}, {Leconte}, {Rouan}, {Forget},
  {L{\'e}ger}, \& {Schneider}}]{2014A&A...563A.103S}
{Samuel}, B., {Leconte}, J., {Rouan}, D., {Forget}, F., {L{\'e}ger}, A., \&
  {Schneider}, J. 2014, \aap, 563, A103

\bibitem[{{Schlawin} {et~al.}(2017){Schlawin}, {Rieke}, {Leisenring}, {Walker},
  {Fraine}, {Kelly}, {Misselt}, {Greene}, {Line}, {Lewis}, \&
  {Stansberry}}]{2017PASP..129a5001S}
{Schlawin}, E., {et~al.} 2017, \pasp, 129, 015001

\bibitem[{Shannon(1948)}]{shannon48}
Shannon, C. 1948, Bell System Technical Journal, 27, 379

\bibitem[{{Sharp} \& {Burrows}(2007)}]{2007ApJS..168..140S}
{Sharp}, C.~M., \& {Burrows}, A. 2007, \apjs, 168, 140

\bibitem[{{Sing} {et~al.}(2009){Sing}, {D{\'e}sert}, {Lecavelier Des Etangs},
  {Ballester}, {Vidal-Madjar}, {Parmentier}, {Hebrard}, \&
  {Henry}}]{2009A&A...505..891S}
{Sing}, D.~K., {D{\'e}sert}, J.-M., {Lecavelier Des Etangs}, A., {Ballester},
  G.~E., {Vidal-Madjar}, A., {Parmentier}, V., {Hebrard}, G., \& {Henry}, G.~W.
  2009, \aap, 505, 891

\bibitem[{{Sing} {et~al.}(2015){Sing}, {Wakeford}, {Showman}, {Nikolov},
  {Fortney}, {Burrows}, {Ballester}, {Deming}, {Aigrain}, {D{\'e}sert},
  {Gibson}, {Henry}, {Knutson}, {Lecavelier des Etangs}, {Pont},
  {Vidal-Madjar}, {Williamson}, \& {Wilson}}]{2015MNRAS.446.2428S}
{Sing}, D.~K., {et~al.} 2015, \mnras, 446, 2428

\bibitem[{{Sing} {et~al.}(2016){Sing}, {Fortney}, {Nikolov}, {Wakeford},
  {Kataria}, {Evans}, {Aigrain}, {Ballester}, {Burrows}, {Deming},
  {D{\'e}sert}, {Gibson}, {Henry}, {Huitson}, {Knutson}, {Etangs}, {Pont},
  {Showman}, {Vidal-Madjar}, {Williamson}, \& {Wilson}}]{2016Natur.529...59S}
---. 2016, \nat, 529, 59

\bibitem[{{Spiegel} \& {Burrows}(2010)}]{2010ApJ...722..871S}
{Spiegel}, D.~S., \& {Burrows}, A. 2010, \apj, 722, 871

\bibitem[{{Stevenson} {et~al.}(2016{\natexlab{a}}){Stevenson}, {Bean},
  {Seifahrt}, {Gilbert}, {Line}, {D{\'e}sert}, \&
  {Fortney}}]{2016ApJ...817..141S}
{Stevenson}, K.~B., {Bean}, J.~L., {Seifahrt}, A., {Gilbert}, G.~J., {Line},
  M.~R., {D{\'e}sert}, J.-M., \& {Fortney}, J.~J. 2016{\natexlab{a}}, \apj,
  817, 141

\bibitem[{{Stevenson} {et~al.}(2016{\natexlab{b}}){Stevenson}, {Lewis}, {Bean},
  {Beichman}, {Fraine}, {Kilpatrick}, {Krick}, {Lothringer}, {Mandell},
  {Valenti}, {Agol}, {Angerhausen}, {Barstow}, {Birkmann}, {Burrows},
  {Charbonneau}, {Cowan}, {Crouzet}, {Cubillos}, {Curry}, {Dalba}, {de Wit},
  {Deming}, {D{\'e}sert}, {Doyon}, {Dragomir}, {Ehrenreich}, {Fortney},
  {Garc{\'{\i}}a Mu{\~n}oz}, {Gibson}, {Gizis}, {Greene}, {Harrington}, {Heng},
  {Kataria}, {Kempton}, {Knutson}, {Kreidberg}, {Lafreni{\`e}re}, {Lagage},
  {Line}, {Lopez-Morales}, {Madhusudhan}, {Morley}, {Rocchetto}, {Schlawin},
  {Shkolnik}, {Shporer}, {Sing}, {Todorov}, {Tucker}, \&
  {Wakeford}}]{2016PASP..128i4401S}
{Stevenson}, K.~B., {et~al.} 2016{\natexlab{b}}, \pasp, 128, 094401

\bibitem[{{Swain} {et~al.}(2008){Swain}, {Vasisht}, \&
  {Tinetti}}]{2008Natur.452..329S}
{Swain}, M.~R., {Vasisht}, G., \& {Tinetti}, G. 2008, \nat, 452, 329

\bibitem[{{Swinyard} {et~al.}(2004){Swinyard}, {Rieke}, {Ressler}, {Glasse},
  {Wright}, {Ferlet}, \& {Wells}}]{2004SPIE.5487..785S}
{Swinyard}, B.~M., {Rieke}, G.~H., {Ressler}, M., {Glasse}, A., {Wright},
  G.~S., {Ferlet}, M., \& {Wells}, M. 2004, in \procspie, Vol. 5487, Optical,
  Infrared, and Millimeter Space Telescopes, ed. J.~C. {Mather}, 785--793

\bibitem[{{Thorngren} {et~al.}(2016){Thorngren}, {Fortney}, {Murray-Clay}, \&
  {Lopez}}]{2016ApJ...831...64T}
{Thorngren}, D.~P., {Fortney}, J.~J., {Murray-Clay}, R.~A., \& {Lopez}, E.~D.
  2016, \apj, 831, 64

\bibitem[{{Waldmann} {et~al.}(2015{\natexlab{a}}){Waldmann}, {Rocchetto},
  {Tinetti}, {Barton}, {Yurchenko}, \& {Tennyson}}]{2015ApJ...813...13W}
{Waldmann}, I.~P., {Rocchetto}, M., {Tinetti}, G., {Barton}, E.~J.,
  {Yurchenko}, S.~N., \& {Tennyson}, J. 2015{\natexlab{a}}, \apj, 813, 13

\bibitem[{{Waldmann} {et~al.}(2013){Waldmann}, {Tinetti}, {Deroo}, {Hollis},
  {Yurchenko}, \& {Tennyson}}]{2013ApJ...766....7W}
{Waldmann}, I.~P., {Tinetti}, G., {Deroo}, P., {Hollis}, M.~D.~J., {Yurchenko},
  S.~N., \& {Tennyson}, J. 2013, \apj, 766, 7

\bibitem[{{Waldmann} {et~al.}(2015{\natexlab{b}}){Waldmann}, {Tinetti},
  {Rocchetto}, {Barton}, {Yurchenko}, \& {Tennyson}}]{2015ApJ...802..107W}
{Waldmann}, I.~P., {Tinetti}, G., {Rocchetto}, M., {Barton}, E.~J.,
  {Yurchenko}, S.~N., \& {Tennyson}, J. 2015{\natexlab{b}}, \apj, 802, 107

\bibitem[{{Wells} {et~al.}(2015){Wells}, {Pel}, {Glasse}, {Wright},
  {Aitink-Kroes}, {Azzollini}, {Beard}, {Brandl}, {Gallie}, {Geers}, {Glauser},
  {Hastings}, {Henning}, {Jager}, {Justtanont}, {Kruizinga}, {Lahuis}, {Lee},
  {Martinez-Delgado}, {Mart{\'{\i}}nez-Galarza}, {Meijers}, {Morrison},
  {M{\"u}ller}, {Nakos}, {O'Sullivan}, {Oudenhuysen}, {Parr-Burman}, {Pauwels},
  {Rohloff}, {Schmalzl}, {Sykes}, {Thelen}, {van Dishoeck}, {Vandenbussche},
  {Venema}, {Visser}, {Waters}, \& {Wright}}]{2015PASP..127..646W}
{Wells}, M., {et~al.} 2015, \pasp, 127, 646

\bibitem[{{West} {et~al.}(2004){West}, {Baines}, {Friedson}, {Banfield},
  {Ragent}, \& {Taylor}}]{2004jpsm.book...79W}
{West}, R.~A., {Baines}, K.~H., {Friedson}, A.~J., {Banfield}, D., {Ragent},
  B., \& {Taylor}, F.~W. 2004, {Jovian clouds and haze}, ed. F.~{Bagenal},
  T.~E. {Dowling}, \& W.~B. {McKinnon}, 79--104

\bibitem[{{Wong} {et~al.}(2004){Wong}, {Mahaffy}, {Atreya}, {Niemann}, \&
  {Owen}}]{2004Icar..171..153W}
{Wong}, M.~H., {Mahaffy}, P.~R., {Atreya}, S.~K., {Niemann}, H.~B., \& {Owen},
  T.~C. 2004, Icarus, 171, 153

\end{thebibliography}

% % % % % % % % % % % % % % % % % % % % % % % % % % % % % % % % % % % % % % % % % % % % % % % % % % % % % % % % % % % % % % % % % % % % % %

\end{document}